\documentclass[apj]{emulateapj}
\usepackage{amsmath,amssymb,latexsym,mathrsfs}
\submitted{}

\begin{document}
\title{The intrinsic abundance ratio and X-factor of CO isotopologues \\in L\,1551 shielded from FUV photodissociation}

\author{Sheng-Jun Lin\altaffilmark{1}, Yoshito Shimajiri\altaffilmark{2}, Chihomi Hara\altaffilmark{3}, Shih-Ping Lai\altaffilmark{1}, Fumitaka Nakamura\altaffilmark{4}, Koji Sugitani\altaffilmark{5}, Ryohei Kawabe\altaffilmark{3,4,6}, Yoshimi Kitamura\altaffilmark{7}, Atsushi Yoshida\altaffilmark{8}, Hidefumi Tatei\altaffilmark{8}, Toshiya Akashi\altaffilmark{8}, Aya E. Higuchi\altaffilmark{10}, and Takashi Tsukagoshi\altaffilmark{9}}
\email{sj.lin@gapp.nthu.edu.tw, slai@phys.nthu.edu.tw}

\altaffiltext{1}{Institute of Astronomy and Department of Physics, National Tsing Hua University, Hsinchu 30013, Taiwan}
\altaffiltext{2}{Laboratoire AIM, CEA/DSM-CNRS-Universit${\rm \acute{e}}$ Paris Diderot, IRFU/Service d'Astrophysique, CEA Saclay, F-91191 Gif-sur-Yvette, France}
\altaffiltext{3}{The University of Tokyo, 7-3-1 Hongo Bunkyo, Tokyo 113-0033, Japan}
\altaffiltext{4}{National Astronomical Observatory of Japan, 2-21-1 Osawa, Mitaka, Tokyo 181-8588, Japan}
\altaffiltext{5}{Graduate School of Natural Sciences, Nagoya City University, Mizuho-ku, Nagoya 467-8501, Japan}
\altaffiltext{6}{SOKENDAI (The Graduate University for Advanced Studies), 2-21-1 Osawa, Mitaka, Tokyo 181-8588, Japan}
\altaffiltext{7}{Institute of Space and Astronautical Science, Japan Aerospace Exploration Agency, 3-1-1 Yoshinodai, Chuo-ku, Sagamihara, Kanagawa 252-5210, Japan}
\altaffiltext{8}{Department of Earth and Planetary Sciences, Tokyo Institute of Technology, 2-12-1, Okayama, Meguro-ku, Tokyo 152-8551, Japan}
\altaffiltext{9}{College of Science, Ibaraki University, 2-1-1 Bunkyo, Mito, Ibaraki 310-8512, Japan}
	
\begin{abstract}
	
We investigate the intrinsic abundance ratio of $^{13}$CO to C$^{18}$O and the X-factor 
in L\,1551 using the Nobeyama Radio Observatory (NRO) 45\,m telescope. 
L\,1551 is chosen because it is relatively isolated in the Taurus molecular cloud shielded 
from FUV photons, providing an ideal environment for studying the target properties.
Our observations cover $\sim$40\arcmin$\times$40\arcmin with resolution $\sim$30\arcsec, 
which are the maps with highest spatial dynamical range to date.
We derive the $X_{\rm ^{13}CO}/X_{\rm C^{18}O}$ value on the sub-parsec scales in the range of $\sim$3--27 
with a mean value of 8.0$\pm$2.8. 
Comparing to the visual extinction map derived from the $\it Herschel$ observations, 
we found that the abundance ratio reaches its maximum at low $A_{\rm V}$ (i.e., $A_{\rm V} \sim$ 1--4\,mag), 
and decreases to the typical solar system value of 5.5 inside L\,1551 MC. 
The high $X_{\rm ^{13}CO}/X_{\rm C^{18}O}$ value at the boundary of the cloud 
is most likely due to the selective FUV photodissociation of C$^{18}$O. 
This is in contrast with Orion-A where its internal OB stars keep the abundance ratio at a high level greater than $\sim$10. 
In addition, we explore the variation of the X-factor, because it is an uncertain but widely used quantity in extragalactic studies. 
We found that X-factor $\propto$ $N_{\rm H_2}^{1.0}$ which is consistent with previous simulations. 
Excluding the high density region, the average X-factor is similar to the Milky Way average value.

\end{abstract}


\section{Introduction}
	
The ultraviolet (UV) radiation plays a crucial role in many processes of the interstellar medium (ISM), such as photoelectric heating, grain charging, photoionization, and photo-dissociation of molecules \citep{bet07}, in which the far-ultraviolet (FUV: 6\,eV $<h\nu<$ 13.6\,eV) radiation from massive stars or interstellar radiation field (ISRF) influences the structure, chemistry, thermal balance, and evolution of neutral interstellar medium \citep{hol97}. 
Therefore, studying these influence helps to understand the process of star formation. 
The FUV emission selectively dissociates CO rare isotopologues more effectively than CO owing to different levels of self-shielding effects \citep{dis88, war96, lis07, vis09, shi14, shi15}. 
For the FUV emission with energy high enough to photodissociate $^{12}$CO, it rapidly becomes optically thick when it penetrates into molecular clouds.
In contrast, the self-shielding effect of C$^{18}$O is relatively less significant because of the shift of its absorption lines and its low abundance. 
Therefore, the FUV emission with energy above the C$^{18}$O dissociation level is expected to penetrate relatively deeper in a molecular cloud, and the abundance ratio of $^{13}$CO and C$^{18}$O, $X_{\rm ^{13}CO}/X_{\rm C^{18}O}$, will increase when the self-shielding effect of C$^{18}$O does not yet dominate.     
When the self-shielding effect of both $^{13}$CO and C$^{18}$O become important, the abundance ratio should decrease toward the intrinsic abundance value which derived from abundances of the elements, $^{12}$C, $^{13}$C, $^{16}$O, and $^{18}$C.
The intrinsic value may vary as the distance to the Galactic center \citep{wil99}.
	
The $X_{\rm ^{13}CO}/X_{\rm C^{18}O}$ has been observed in regions with various conditions. 
In the typical massive star-forming regions, where the ISM is filled up with diffuse FUV flux from OB stars, the measurements of $X_{\rm ^{13}CO}/X_{\rm C^{18}O}$ are usually higher than the intrinsic value.
\citet{shi14} measured $X_{\rm ^{13}CO}/X_{\rm C^{18}O}$ in the Orion-A giant molecular cloud (Orion-A GMC) and found that the $X_{\rm ^{13}CO}/X_{\rm C^{18}O}$ of most regions are a factor of two greater than 5.5, the typical value in the solar system.
It is possible that besides the interstellar FUV radiation, embedded OB stars provide strong FUV radiation so that  the distance of the penetration will increase.  
Another possibility is that the higher temperature in the massive cores will also change the fractionation of $^{12}$C and $^{13}$C \citep{rol13}. 
In the intermediate-mass star-forming regions, 
\citet{kon15} showed that the intensity ratio of $^{13}$CO ($J$=2--1) to C$^{18}$O ($J$=2--1), which is equivalent to $X_{\rm ^{13}CO}/X_{\rm C^{18}O}$ for optically thin case, rises to a peak up to $\sim$40 at $A_{\rm V} \sim$ 5\,mag then decreases to $\sim$4.5 with increasing $A_{\rm V}$ in the southeast of the California molecular cloud. Although the trend is consistent with the theoretical expectations, the peak occurs at somewhat higher extinction; \citet{war96} showed that the peak is at $A_{\rm V} \sim$1--3\,mag for regions with various density from 10$^2$ to 10$^5$\,cm$^{-3}$. 
In the low-mass star-forming regions, \citet{lad94} observed a part of the IC 5146 filament and found that the $X_{\rm ^{13}CO}/X_{\rm C^{18}O}$ value is considerably greater than 5.5 in the outer parts ($A_{\rm V} \lesssim$ 10\,mag). 

LDN 1551 (hereafter L\,1551) is a relatively isolated nearby star-forming region located at a distance of 160\,pc \citep{sne81, ber99} at the south end of the Taurus-Auriga-Perseus molecular cloud complex. Two small clusters of young stellar objects (YSOs) were detected in L\,1551. One contains two embedded Class I sources, L\,1551 IRS 5 (hereafter IRS 5) and L\,1551 NE (hereafter NE), and the other is a group of more evolved YSOs (hereafter HL Tau group) located at the north of IRS 5 and NE. 
The IRS 5/NE cluster co-host a parsec-scale bipolar outflow \citep{sne80} which is likely mixed from outflows of each source \citep{mor06}.
Several Herbig-Haro objects along the outflows have been identified with different origins \citep{dev99}. 
Another east-west redshifted outflow was found later \citep{mor91, pou91}, but since IRS 5 and NE are two binary systems \citep{loo97,rei02,tak14,cho14} the origin of this east-west outflow is still questionable \citep{mor06, sto06}. 
The HL Tau group contains HL Tau, XZ Tau, LkH$\alpha$ 358, and V1213 Tau.
The jets and outflows of HL Tau group were detected in optical observations \citep{mun90, bur96}. In addition to the protostars, a gravitationally bound starless core, L\,1551 MC, is found at the north-western side of the IRS 5/NE cluster which may have started gravitational collapse or could be supported by magnetic fields of $\sim$160\,$\mu$G \citep{swi05, swi06}. 
Since L\,1551 dose not contain OB stars, L\,1551 is a suitable target for studying the FUV influence only from the interstellar radiation.

In this paper, we aim to study the variation of $X_{\rm ^{13}CO}/X_{\rm C^{18}O}$ under the influence of the interstellar FUV radiation. 
In contrast to previous studies, we observed $^{12}$CO ($J$=1--0), $^{13}$CO ($J$=1--0), and C$^{18}$O ($J$=1--0) lines using the 25-BEam Array Receiver System (BEARS) receiver equipped on the Nobeyama Radio Observatory (NRO) 45\,m telescope to obtain data that could resolve the sub-parsec scale with a complete coverage from the outskirts of the cloud into its dense region. 
We also use $\it Herschel$ archival data to obtain a visual extinction map up to $A_{\rm V} \sim$70\,mag in order to examine the variation of $X_{\rm ^{13}CO}/X_{\rm C^{18}O}$ with the FUV attenuation.
In \S \ref{sec_obs}, we describe our NRO observations and Herschel data.  
In \S \ref{sec_result}, 
we present the $^{12}$CO, $^{13}$CO, and C$^{18}$O maps of L\,1551, and then derive the excitation temperature and optical depths of the $^{13}$CO and C$^{18}$O lines and the column densities of these molecules. 
In \S \ref{sec_discussion}, we discuss the dependence of the abundance ratio of $^{13}$CO to C$^{18}$O on visual extinction, $A_{\rm V}$, the influence of FUV radiation, and the X-factor in L\,1551.
At the end, we summarize our results in \S \ref{sec_summary}.

\section{Observations and data reduction}\label{sec_obs}

\subsection{NRO 45m observations}

We observed the L\,1551 star-forming region using the 45\,m telescope at NRO. Three molecular lines were observed: $^{12}$CO ($J$=1--0; 115.27120\,GHz), $^{13}$CO ($J$=1--0; 110.20135\,GHz), and C$^{18}$O ($J$=1--0; 109.78218\,GHz).  The $^{12}$CO data has been published in \citet{yos10}. 
The observations were carried out between 2007 to 2010 (Table 1).
The telescope has beam sizes (HPBW), $\theta_{\rm HPBW}$, of 15$\arcsec$ at 115\,GHz, 
and of 16$\arcsec$ at 110\,GHz. 
The main beam efficiencies, $\eta_{\rm mb}$, were measured as 32$\%$ in 2007--2008 season at 115\,GHz, 38$\%$ in 2009--2010 season at 110\,GHz, and  43$\%$ in 2009 season at 110\,GHz. 
These efficiencies were measured every season by NRO staff with a superconductor-insulator-superconductor (SIS) receiver, S100,
and an acousto-optical spectrometers (AOSs). 
The front end is BEARS, which has 25 beams configured in a 5-by-5 array with a beam separation of 41$\farcs$1 \citep{sun00}. 
BEARS is operated under double-sideband (DSB) mode with each beam connected to a set of 1024 channel auto-correlators (ACs) as the back end.
We set each set of ACs to a bandwidth of 32\,MHz and a resolution of 37.8\,kHz \citep{sor00}, corresponding to velocity resolutions of $\sim$0.1\,km\,s$^{-1}$ at the rest frequencies of the three lines. 
We calibrated the variations in both beam efficiency and sideband ratio of the 25 beams with the values from the NRO website.  
These values were measured by NRO staff every season through measurements of bright sources with S100 in SSB mode.

For the observations, we used the on-the-fly (OTF) mapping technique \citep{saw08}. 
The antenna is driven at a constant speed to continuously scan our observing region toward L\,1551. 
To get rid of artificial scanning patterns in the results, we scanned the observed regions twice, one in RA and one in Dec directions, and then combined the OTF maps in the two orthogonal directions by PLAIT algorithm \citep{emr88}. 
The antenna pointing was checked every $\sim$60--70 minutes by observing the SiO maser source NML-Tau, and the pointing accuracy was about 3$\arcsec$ during the observations. Figure \ref{fig_intro} (a), (b), and (c) show the observed areas of $^{12}$CO, $^{13}$CO, and C$^{18}$O, covering 44$\arcmin$ $\times$ 44$\arcmin$, 42$\arcmin$ $\times$ 43$\arcmin$, and 30$\arcmin$ $\times$ 30$\arcmin$, respectively.

The data were converted in terms of the main-beam brightness temperature in units of K, $T_{\rm mb}=T_{\rm A}^*/\eta_{\rm mb}$, where $T_{\rm A}^*$ is the antenna temperature in units of K. In order to maximize the energy concentration ratio, we applied a spheroidal function with $m = 6$ and $\alpha=1$ \citep{sch84} to convolve the OTF data, and then obtained three dimensional $(\alpha,\,\delta,\,v_{\rm LSR})$ final cube data.  
Adopting Nyquist sampling for the 45\,m telescope, we set the spatial grid size to 7$\farcs$5, and the final effective beam sizes, $\theta_{\rm eff}$, are 21$\farcs$8 for $^{12}$CO and 22$\farcs$2 for $^{13}$CO and C$^{18}$O. 
The rms noise 1$\sigma$ levels are 1.23\,K for $^{12}$CO, 0.94\,K for $^{13}$CO, and 0.67\,K for C$^{18}$O in $T_{\rm mb}$ for a velocity resolution of 0.1\,km\,s$^{-1}$. 
To achieve higher signal-to-noise ratios, we further convolved all maps with Gaussians to make channel maps, integrated intensity maps, mean velocity maps, and velocity dispersion maps, and the effective beam size of each map is written in its figure caption.
We summarize the parameters for each line in Table \ref{obs_parameter}.

\subsection{{\it Herschel} column density and dust temperature maps}\label{sec_obs_her}

We used the  Herschel Science Archival SPIRE/PACS image data (Observation ID: 1342202250/1342202251, Quality: level 2 processed) of 160, 250, 350, and 500 $\mu$m to make dust temperature ($T_{\rm dust}$) and column density maps in a similar way to \citet{kon10}. 
Since the 70\,$\mu$m emission seems to be not detected except towards IRS 5/NE and the HL Tau group, we used only 4 bands (500, 350, 250, and 160\,$\mu$m).
	
At first, we made convolutions for all $\it Herschel$ images (other than the 500\,$\mu$m image) to smooth their resolutions to the 500\,$\mu$m resolution of 36$\arcsec$ by using the IDL package developed by \citet{ani11}. 
Then, we resampled up or down all the images (other than the 250\,$\mu$m) to the same grid size of the 250\,$\mu$m image (6$\arcsec$) and derived a spectral energy distribution (SED) at each position of the 250\,$\mu$m image. 
Here, we adopted an area of $\sim3\arcmin\times3 \arcmin$, centered at (RA$_{\rm J2000}$, Dec$_{\rm J2000}$) = (4$^h$30$^m$34$\fs$9, +18$\arcdeg$28$\arcmin$55$\arcsec$), as an area of the zero point of the surface brightness for the L\,1551 cloud.
	
Assuming a single temperature of the dust emission, the gray-body SED fitting was performed with a function of 
\begin{equation}
I_\nu = B_\nu (T_{\rm dust}) (1-e^{\tau_\nu}),
\end{equation}
where $\nu$ infers the frequency, $B_\nu$ expresses Planck's law, $I_\nu$ is the observed surface brightness, and $\tau_\nu$ is the dust optical depth.  $\tau_\nu$ can be expressed as $\kappa_\nu \Sigma$, where $\kappa_\nu$ is the dust opacity per unit mass and $\Sigma$ is the surface mass density. 
We adopted 
\begin{equation}\label{equ_k}
\kappa_\nu = 0.1 (\frac{\nu}{1000\,{\rm GHz}})^\beta\,\rm cm^2\,g^{-1}, 
\end{equation}
where $\beta$ = 2, following \citet{kon10}.
	
We carried out the SED fitting at each point where the dust emission was detected above 3 times of the rms noise level, which was measured in the area of the zero point of the surface brightness at all these 4 bands. 
In the case that the SED fitting failed or the surface brightness was not high enough for SED fitting at some points, we set NaN there. 
As the data weight of the SED fitting, we adopted $1/\sigma^2$, where $\sigma^2$ is the square sum of the rms noise and the calibration uncertainties of surface brightness (15$\%$ at 500, 350, and 250\,$\mu$m from \citet{gri10}; 20$\%$ at 160\,$\mu$m from \citet{pog10}). 
The dust temperature of each pixel was derived from the above SED fitting, and the result is shown in Fig.\,\ref{fig_intro} (f). 
We used $\chi^2$ statistic to determine the goodness of the SED fitting per pixel \citep{pre07}. For the SED fitting, each $\chi^2$ statistic has a $\chi^2$ distribution with two degrees of freedom, and then the probability Q gives a quantitative measure for the goodness. In most of the inner region of the L\,1551 cloud, the SED fitting seems good (the probability Q $\sim$ 0.8), while in the region around YSOs and toward the outskirts, the fitting seems bad with small probability Q. The badness around YSOs may be due to multiple-temperature components presenting in the SED.
	
To obtain a column density map with a higher resolution of 18$\arcsec$ (Fig.\,\ref{fig_intro} (d)), we derived the column density directly from the 250\,$\mu$m surface brightness ($\tau_{\rm 250\,\mu m}\ll 1$ from previous fitting) adopting the dust temperature of the SED fitting with a lower resolution of 36$\arcsec$ \citep{kon10}.
	
\section{Results}\label{sec_result}

\subsection{$^{12}$CO ($J$=1--0) emission line}

Figure \ref{fig_intro} (a) shows the integrated intensity map of $^{12}$CO ($J$=1--0) integrated in the $v_{\rm LSR}$ range from $-$15\,km\,s$^{-1}$ to 20\,km\,s$^{-1}$ for pixels with its signal-to-noise ratio greater than 3. 
The $^{12}$CO emission is distributed all over the observed area. 
A sharp edge at the southeast of the mapping area can be recognized. \citet{mor06} suggested that the stellar winds from massive stars, Betelgeuse and Rigel in the Ori-Eridanus supershell, possibly compressed the edge of the L\,1551 molecular cloud. 
Two elongated structures which are extended in the southwest--northeast direction and centered on the Class I source IRS 5 are identified as the molecular outflows ejected from IRS 5 and another embedded Class I source NE \citep{sne80,mor06,sto06}.

\subsection{$^{13}$CO ($J$=1--0) emission line}

Figure \ref{fig_intro} (b) shows the integrated intensity map of $^{13}$CO ($J$=1--0) integrated in the $v_{\rm LSR}$ range from $-$15\,km\,s$^{-1}$ to 20\,km\,s$^{-1}$ for pixels with its signal-to-noise ratio greater than 3. The overall distribution of $^{13}$CO is consistent with those in previous studies of $^{13}$CO carried out with position switching mode of the FCRAO 14\,m \citep{sto06} and NRO 45\,m \citep{yos10} telescopes.
There are a cavity structure at the northeast of NE and a U-shaped wall structure at the southwest of IRS 5. 
In addition, two intensity peaks can be seen toward IRS 5 and NE. 
These structures are not recognized in the $^{12}$CO maps.

\subsection{C$^{18}$O ($J$=1--0) emission line}

Figure \ref{fig_intro} (c) shows the integrated intensity map of C$^{18}$O ($J$=1--0) integrated in the $v_{\rm LSR}$ range from $-$15\,km\,s$^{-1}$ to 20\,km\,s$^{-1}$ for pixels with its signal-to-noise ratio greater than 3. The C$^{18}$O emission line is likely to trace the inner part of the regions traced by $^{12}$CO and $^{13}$CO. The overall distribution of C$^{18}$O is consistent with those in previous studies of C$^{18}$O carried out with the OTF mode of the Kitt Peak 12\,m telescope \citep{swi05} and the position switching mode of the NRO 45\,m \citep{yos10} telescope.

\subsection{Column densities of the $^{13}$CO and C$^{18}$O gas and abundance ratio of $^{13}$CO to C$^{18}$O}\label{result_col_abu}

In order to derive the optical depths and column densities of $^{13}$CO and C$^{18}$O, 
we assume that (1) $^{12}$CO ($J$=1--0) is optically thick, 
(2) $^{12}$CO and its rarer isotopic species trace the same component, and 
(3) these three lines reach Local Thermal Equilibrium (LTE). 
Thus, the temperature of $^{12}$CO ($J$=1--0) can be treated as the excitation temperature, $T_{\rm ex}$, of $^{13}$CO ($J$=1--0) and C$^{18}$O ($J$=1--0) for deriving the optical depths and column densities of $^{13}$CO and C$^{18}$O. 
In order to make direct comparison between the different lines, we smooth all the data so that they have the same effective beam  size of 30$\farcs$4 (corresponding to 0.023\,pc at the distance of 160\,pc). 
Then, we obtain their peak intensities and full widths at half maximum (FWHM) by applying Gaussian fitting to spectra of the $^{12}$CO ($J$=1--0), $^{13}$CO ($J$=1--0), and C$^{18}$O ($J$=1--0) cube data for pixels with their signal-to-noise ratio greater than 5.   

In general, the spectra of $^{13}$CO ($J$=1--0) and C$^{18}$O ($J$=1--0) show single-component velocity structures, 
and thus we can apply  single-Gaussian fitting as an appropriate method. 
However, since the spectra of $^{12}$CO ($J$=1--0) often show multiple velocity components due to the prominent outflows, 
we adopt different fitting strategies to obtain the excitation temperature. 
First, if the $^{12}$CO spectrum has only one velocity component, 
which usually happen in quiescent ambient region, the single-Gaussian fitting may gives an adequate result. 
If the residual of the fitting at the peak velocity is less than 3$\sigma$,
we take the peak intensity to calculate the excitation temperature.
Second, if the single-Gaussian fitting fails, we apply double-Gaussian fitting to separate the ambient and outflow components. 
Except the regions around NE, only one redshifted or blueshifted component appears. 
We identify the component of which the peak velocity is similar to the peak velocities of $^{13}$CO and C$^{18}$O spectra as the ambient component and the other component as the outflow component, 
if (1) the residuals of the fitting at both the peak velocities are less than 3$\sigma$ 
and (2) the velocity ranges of the two Gaussian components at half maximum are not overlapped with each other. 
The later criterion is included, 
because the determination of the peak intensity of the ambient component could be affected by the close outflow component.   
In some cases, the peak $^{12}$CO intensity is even smaller than the peak $^{13}$CO intensity, which is not physical and is rejected. 
Third, for the remaining spectra, 
the velocity structure seems to contain three or more components.  
Since our goal is to obtain the peak intensity of the ambient gas, 
we only apply a single Gaussian to one peak having the peak velocity closest to the ambient gas velocity determined from the $^{13}$CO and C$^{18}$O spectra at the same position.
We fix the peak velocity of this single Gaussian as the mean value of the $^{13}$CO and C$^{18}$O peak velocities 
and we only fit the Gaussian within a narrow velocity range bracketed by the two local minima around the peak.
Following the above procedure, we obtain the estimate of the excitation temperature from the peak intensity of the fitted Gaussian to the ambient $^{12}$CO emission. 

\begin{figure}
	\includegraphics[scale=.7]{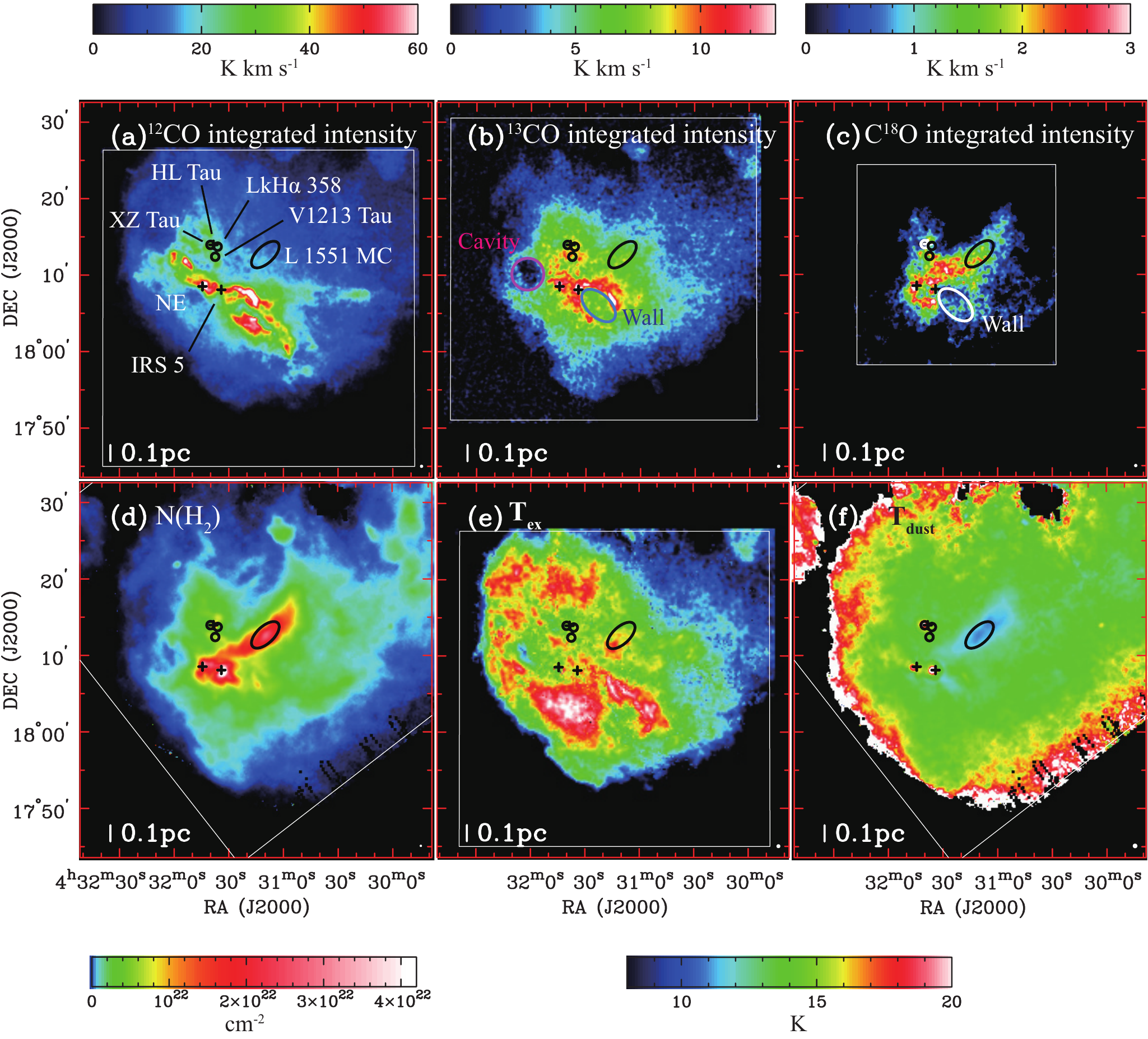}
	\caption{L\,1551 maps of integrated intensity and derived properties. The integrated intensity maps of (a) $^{12}$CO ($J$=1--0), (b) $^{13}$CO ($J$=1--0), and (c)  C$^{18}$O ($J$=1--0) in $v_{\rm LSR}$=[$-$15\,km\,s$^{-1}$, 20\,km\,s$^{-1}$] in units of K km\,s$^{-1}$ with effective beam sizes of 24$\farcs$9, 25$\farcs$2, and 25$\farcs$2, respectively. The crosses denote the Class I sources NE, and IRS 5, and the Class I/II source HL Tau. The circles denote XZ Tau, LkH$\alpha$ 358 and V1213 Tau. The black ellipse denotes the starless core L\,1551 MC. The blue and white ellipses denote a U-shaped wall structure and the magenta one a cavity structure. The linear scale of 0.1\,pc is shown at the bottom-left corner of each panel. (d) The column density map of H$_2$ in units of cm$^{-2}$ with an effective beam size of 18$\arcsec$, derived from {\it Herschel} data. (e) The excitation temperature map in units of K with an effective beam size of 30$\farcs$4, derived from the $^{12}$CO ($J$=1--0) peak intensities. (f) The dust temperature map in units of K with an effective beam size of 36$\arcsec$. Each beam size is denoted as the white ellipses at the bottom-right corner of each panel.}
	\label{fig_intro}
\end{figure}

Figure \ref{fig_intro} (e) shows the excitation temperature map deriving from the following equation with the previous assumption (1) and the beam filling factor of 1 \citep[e.g.,][]{pin10,kon15}
\begin{equation}\label{equ_Tex}
\begin{split}
T_{\rm ex} &= \frac{h\nu_{\rm ^{12}CO}}{k} \left[\log \left(1+\frac{h\nu_{\rm ^{12}CO}/k}{T_{\rm mb,^{12}CO}+J_{\nu}(T_{\rm bg})}\right)\right]^{-1}\rm\,K\\
&= 5.53 \left[\log\left(1+\frac{5.53}{T_{\rm mb,^{12}CO}+0.818}\right)\right]^{-1} \rm\,K,
\end{split}
\end{equation}
where $T_{\rm mb,^{12}CO}$ is the peak intensity of $^{12}$CO ($J$=1--0) in units of K from the above fitting,
$J_{\nu}(T)=\frac{h\nu/k}{\exp(h\nu/(kT))-1}$ is the effective radiation temperature \citep{uli76}, and $T_{\rm bg}$=2.7\,K is the temperature of cosmic microwave background radiation. 
Hereafter, we call this excitation temperature derived from Eq.\,(\ref{equ_Tex}), $T_{\rm CO}$. 
We then obtain the optical depths, $\tau$, of the $^{13}$CO ($J$=1--0) and C$^{18}$O ($J$=1--0) emission and the column densities, $N$, of the $^{13}$CO and C$^{18}$O gas using the following equations \citep[e.g.,][]{lad94, kaw98}
\begin{equation}
\tau_{\rm ^{13}CO}=-\log \left(1-\frac{T_{\rm mb,^{13}CO}/\phi_{\rm ^{13}CO}}{5.29([\exp(5.29/T_{\rm ex})-1]^{-1}-0.164)}\right),
\end{equation}
\begin{equation}
\tau_{\rm C^{18}O}=-\log\left(1-\frac{T_{\rm mb,C^{18}O}/\phi_{\rm C^{18}O}}{5.27([\exp(5.27/T_{\rm ex})-1]^{-1}-0.1666)}\right),
\end{equation}
\begin{equation}
N_{\rm ^{13}CO}=2.42\times 10^{14}\,\frac{\tau_{\rm ^{13}CO}{\rm FWHM}_{\rm ^{13}CO}T_{\rm ex}}{1-\exp(-5.29/T_{\rm ex})}\, \rm cm^{-2},
\end{equation}
and
\begin{equation}
N_{\rm C^{18}O}=2.42\times 10^{14}\,\frac{\tau_{\rm C^{18}O}{\rm FWHM}_{\rm C^{18}O}T_{\rm ex}}{1-\exp(-5.27/T_{\rm ex})} \, \rm cm^{-2},
\end{equation}
where $T_{\rm mb,^{13}CO}$ and $T_{\rm mb,C^{18}O}$ are the peak intensities in K, 
$\rm FWHM_{\rm ^{13}CO}$ and $\rm FWHM_{\rm C^{18}O}$ are the FWHMs in km\,s$^{-1}$, 
and $\phi_{\rm ^{13}CO}$ and $\phi_{\rm C^{18}O}$ are the beam filling factors. 
The beam filling factors can be expressed as $\phi=\theta^2_{\rm source}/(\theta^2_{\rm source}+\theta^2_{\rm beam})$, 
where $\theta_{\rm source}$ and $\theta_{\rm beam}$ are the source size and the effective beam size, respectively. 
In molecular clouds, the C$^{18}$O emission is usually considered to trace the dense cores, clumps, and/or filaments.
On the other hand, the $^{13}$CO emission traces more extended regions than C$^{18}$O. 
In the case of the Taurus molecular cloud, 
\citet{tac02} and \citet{qia12} found that both the typical $^{13}$CO and C$^{18}$O core size are $\sim$0.1\,pc.
Since the effective beam sizes we used here are 30$\farcs$4 (corresponding to 0.023\,pc at the distance of 160\,pc) 
which is much smaller than the typical core size, 
we can assume the emission of our sources fill up the beam and the beam filling factors become $\sim$0.95.
Therefore, we simply adopt $\phi_{\rm ^{13}CO}$ and $\phi_{\rm C^{18}O}$ as 1. 

\begin{figure}
	\includegraphics[scale=.7]{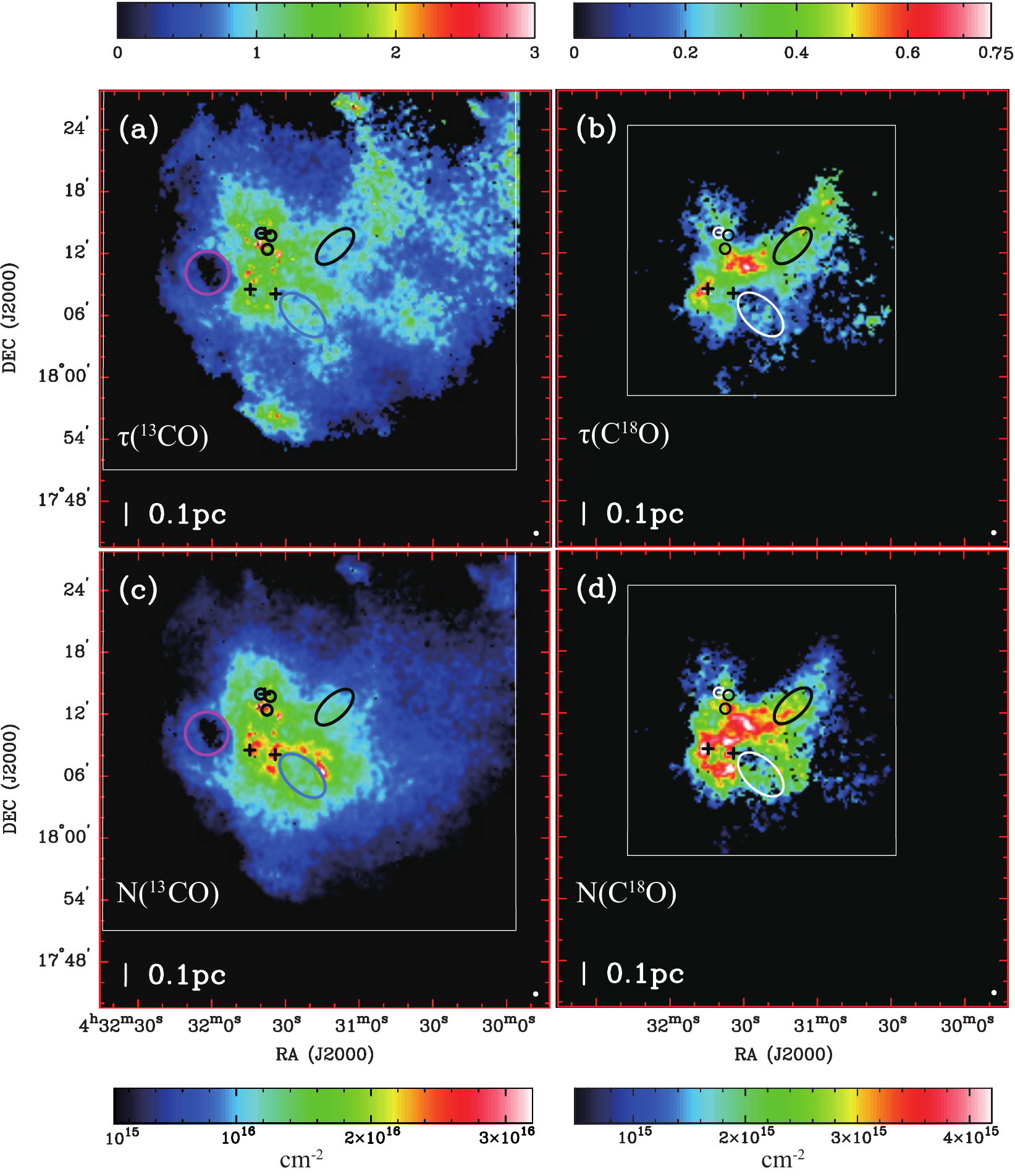}
	\caption{L\,1551 maps of optical depth and column density. The optical depth maps of (a) $^{13}$CO ($J$=1--0) and (b) C$^{18}$O ($J$=1--0) and the column density maps of (c) $^{13}$CO and (d) C$^{18}$O in units of cm$^{-2}$. The effective beam sizes are 30$\farcs$4.\label{fig_tau_N} Note that these maps are made from the data with the signal-to-noise ratio larger than 5. The symbols in the panels have the same meaning as in Fig.\,1.}
\end{figure}

Figures \ref{fig_tau_N} (a) and \ref{fig_tau_N} (b) show the optical depth maps of $^{13}$CO ($J$=1--0) and C$^{18}$O ($J$=1--0),  respectively. 
For the $^{13}$CO ($J$=1--0) emission, its optical depth is more than 1.5 at the center of L\,1551, 
and drops to less than 1 at the outer edge. 
For the C$^{18}$O ($J$=1--0) emission, its optical depth is less than 0.8 in the whole region, 
suggesting that the C$^{18}$O ($J$=1--0) emission is fully optically thin in L\,1551.

Figures \ref{fig_tau_N} (c) and \ref{fig_tau_N} (d) show the column density maps of $^{13}$CO and C$^{18}$O, respectively. 
The cavity structure and the U-shaped wall structure can be seen in the $^{13}$CO column density map. 
The C$^{18}$O column density map shows that C$^{18}$O concentrate in the region surrounded by IRS 5, NE, and L\,1551 MC.
Moreover, the U-shaped wall structure is also seen in the C$^{18}$O column density map as a low column density part.

We can derive the abundances from $X_{\rm ^{13}CO}=N_{\rm ^{13}CO}/N_{\rm H_2}$ and $X_{\rm C^{18}O}=N_{\rm C^{18}O}/N_{\rm H_2}$
with the H$_2$ column density derived from the FIR dust continuum images.
Because we adopt the dust temperature map with an effective beam size of 36$\arcsec$ (see \S \ref{sec_obs_her}), we convolve the $N_{\rm ^{13}CO}$ and $N_{\rm C^{18}O}$ maps with Gaussians to degrade their resolutions down to 36$\arcsec$ (corresponding to 0.027\,pc at the distance of 160\,pc).
Figures \ref{fig_XX} (a) and \ref{fig_XX} (b) show the abundance maps of $^{13}$CO and C$^{18}$O, respectively.
The mean and standard deviation of $X_{\rm ^{13}CO}$ and $X_{\rm C^{18}O}$ are (3.1$\pm$1.2)$\times$10$^{-6}$ and (3.1$\pm$1.2)$\times$10$^{-7}$, respectively. 
We can see $N_{\rm H_2}$ peaks toward IRS 5, NE, and L\,1551 MC in Fig.\,\ref{fig_intro} (d) and $N_{\rm C^{18}O}$ also concentrates in the above regions in Fig.\,\ref{fig_tau_N} (d) but the peaks of $N_{\rm C^{18}O}$ are not as obvious as $N_{\rm H_2}$.
As a result, the $X_{\rm C^{18}O}$ map shows low values in above three regions,
which means the C$^{18}$O depletion occurs at these regions.
It is reasonable for the depletion at L\,1551 MC because we find that L\,1551 MC is dense and cold in Figures \ref{fig_intro} (d) and (f).
For IRS 5 and NE, the dust temperature at the envelop of protostars are higher than other regions,
but the dust temperature is still less than the CO evaporation temperature of 20--25\,K and the H$_2$ column density is very high compared with other regions; therefore the envelop of protostars also show C$^{18}$O depletion.
On the other hand, the low $X_{\rm ^{13}CO}$ value also appears at the C$^{18}$O depletion region; however, the low $X_{\rm ^{13}CO}$ value derived in this region may not show the real situation because $^{13}$CO and dust emission may not trace the exactly same layer in the line of sight (see \S \ref{temperature}).

The abundance ratio of $^{13}$CO to C$^{18}$O can be directly derived from $X_{\rm ^{13}CO}/X_{\rm C^{18}O}$ = $N_{\rm ^{13}CO}/N_{\rm C^{18}O}$.
Figure \ref{fig_XX} (d) shows the spatial variation of $X_{\rm ^{13}CO}/X_{\rm C^{18}O}$. 
The $X_{\rm ^{13}CO}/X_{\rm C^{18}O}$ value ranges from $\sim$3 to $\sim$27 in the whole region, 
and the mean and standard deviation of the abundance ratios are 8.1$\pm$2.8. 
Spatially, the $X_{\rm ^{13}CO}/X_{\rm C^{18}O}$ value in the outskirts of the cloud is more than $\sim$8 
and drops to less than 5.5 around L\,1551 MC 
except for the HL Tau group and the southwest blueshifted outflow lobe with the brightest $^{12}$CO integrated intensity.

\section{Discussion}\label{sec_discussion}

\subsection{Selective photodissociation of C$^{18}$O}\label{discussion_C18O}

\begin{figure}
	\includegraphics[scale=.6]{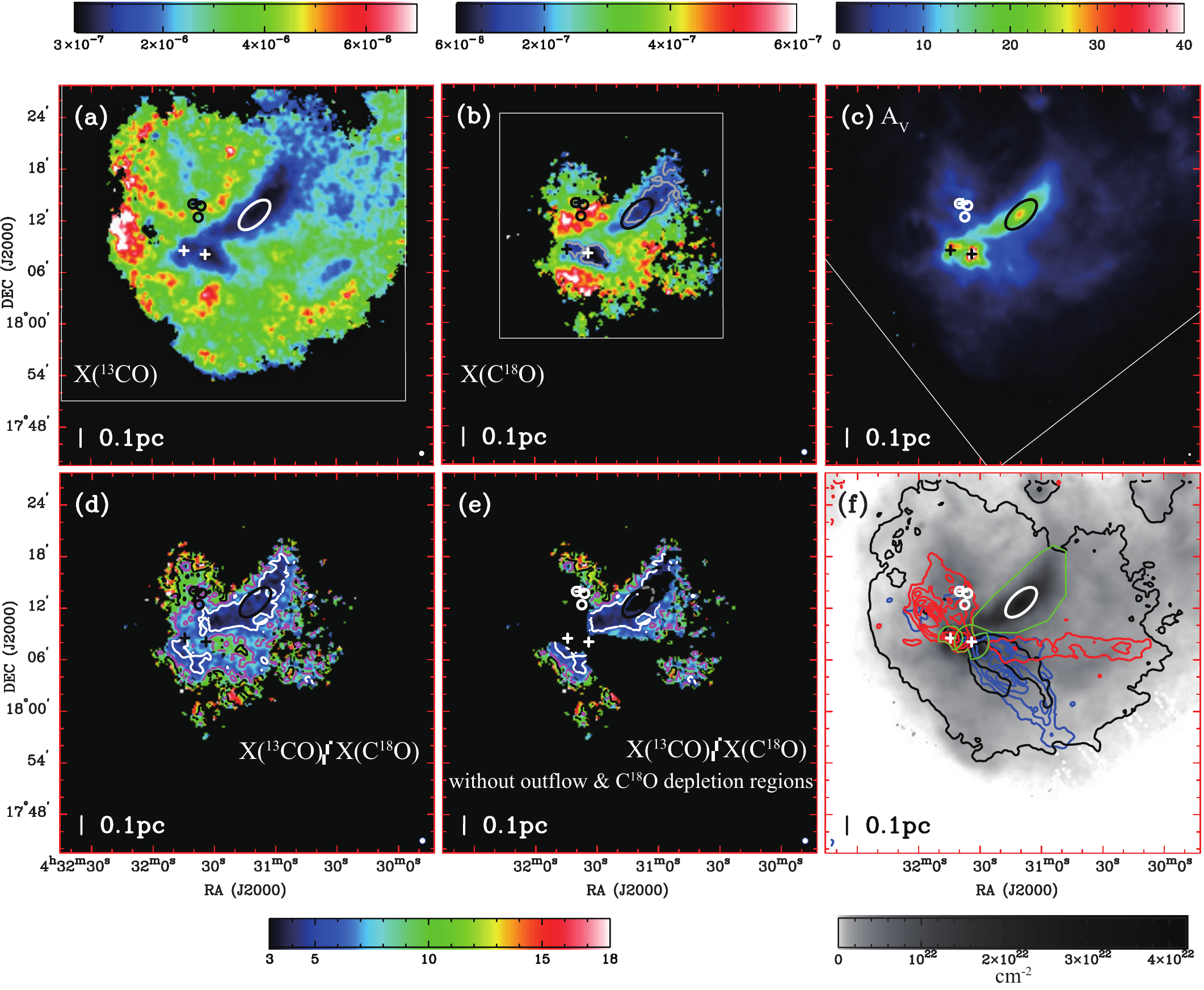}
	\caption{Comparison between the abundances, $X_{\rm ^{13}CO}$ and $X_{\rm C^{18}O}$, the abundance ratio, $X_{\rm ^{13}CO}/X_{\rm C^{18}O}$, and the visual extinction, $A_{\rm V}$. (a) The $X_{\rm ^{13}CO}$ map and (b) the $X_{\rm C^{18}O}$ map with effective beam sizes of 36$\arcsec$. The gray contour in the panel (b) indicates the ISM standard value, $X_{\rm C^{18}O}$ = 1.6$\times$10$^{-7}$. (c) The  $A_{\rm _V}$ map in units of mag, derived from {\it Herschel} data. (d) The $X_{\rm ^{13}CO}/X_{\rm C^{18}O}$ map and (e) the $X_{\rm ^{13}CO}/X_{\rm C^{18}O}$ map without outflow and C$^{18}$O depletion regions. The effective beam sizes are 36$\arcsec$. The white, magenta, and black contours in the panels (d) and (e) indicate 5.5, 8.25, 11\,mag, respectively. (f) The $^{12}$CO ($J$=1--0) integrated intensity contours with an effective beam size of 36$\arcsec$ superposed on the column density map of H$_2$ with an effective beam size of 18$\arcsec$. The contours indicate the blueshifted outflows (blue, $v_{\rm LSR}$ = [$-$15\,km\,s$^{-1}$, 5.5\,km\,s$^{-1}$]), the redshifted outflows (red, $v_{\rm LSR}$ = [7.9\,km\,s$^{-1}$, 20\,km\,s$^{-1}$]), and the ambient component (black, $v_{\rm LSR}$ = [5.5\,km\,s$^{-1}$, 7.9\,km\,s$^{-1}$]). The contour levels start from 3$\sigma$ with a step of 6$\sigma$ where the rms noise 1$\sigma$ levels are 1.44\,K km\,s$^{-1}$ (red), 2.68\,K km\,s$^{-1}$ (black), 1.23\,K km\,s$^{-1}$ (blue). The green polygon is a selected region of L\,1551 MC and the two incomplete green circles are selected regions of IRS 5 and NE, respectively. The symbols in the panels have the same meaning as in Fig.\,1.}\label{fig_XX}
\end{figure}

To study the influence of interstellar FUV radiation on $X_{\rm ^{13}CO}/X_{\rm C^{18}O}$, 
we remove the outflow regions to avoid the influence from outflow activities for later discussion.
We define the outflow regions by the regions where the $^{12}$CO integrated intensities in either blueshifted or redshifted parts are higher than 3$\sigma$ (see Fig.\,\ref{fig_XX} (f)); here we regard $-$15 to 5.5\,km\,s$^{-1}$ and 7.9 to 20\,km\,s$^{-1}$ as the blueshifted and redshifted outflow velocity ranges, respectively. Moreover, to avoid the influence from underestimation of $X_{\rm C^{18}O}$ by C$^{18}$O depletion,
we also remove the region where the $X_{\rm C^{18}O}$ value is less than the ISM standard value of 1.6$\times$10$^{-7}$ \citep{fre82, for11} (shown as the gray contour in Fig. \ref{fig_XX} (b)).
In the other words, we remove the region where C$^{18}$O depletion factor = $1.6\times10^{-7}/X_{\rm C^{18}O}>$ 1.

Figure \ref{fig_XX} (e) shows the abundance map excluding the outflow and C$^{18}$O depletion regions. 
We can see that the spatial variation of the $X_{\rm ^{13}CO}/X_{\rm C^{18}O}$ value is outside-in deceasing and the minimum of the $X_{\rm ^{13}CO}/X_{\rm C^{18}O}$ value coincides with the dense part in L\,1551 (see the H$_2$ column density map in Fig.\,\ref{fig_intro} (d) as a comparison).
This spatial variation might be caused by the penetration capability of interstellar FUV radiation which is related to the visual extinction, $A_{\rm V}$.
In order to investigate the dependence of the abundance ratio on the visual extinction within the molecular cloud, we derived the $A_{\rm V}$ map from the H$_2$ column density map with a relation from \citet{boh78},
\begin{equation}
A_{\rm V}=\frac{N_{\rm H_2}}{9.4\times10^{20}\,\rm cm^{-2}}.\label{equ_Av}
\end{equation}
Figure \ref{fig_XX} (c) shows the derived $A_{\rm V}$ map. 
Since we adopt the dust temperature map with an effective beam size of 36$\arcsec$ (see \S \ref{sec_obs_her}), we convolve the abundance ratio maps with Gaussians to degrade their resolutions down to 36$\arcsec$ (corresponding to 0.027\,pc at the distance of 160\,pc).

Figure \ref{fig_Av_XX} (a) shows the correlation between the $X_{\rm ^{13}CO}/X_{\rm C^{18}O}$ value and the $A_{\rm V}$ value. 
The $X_{\rm ^{13}CO}/X_{\rm C^{18}O}$ value is higher than the typical solar system value of 5.5 in the $A_{\rm V}$ range less than 10\,mag, and then decreases down to $\sim$5.5 with increasing $A_{\rm V}$ value, suggesting the dependence of the abundance ratio on $A_{\rm V}$ value is significant. 
Although the sample of $A_{\rm V} \lesssim$ 2\,mag is sparse because we derived the abundance ratio from the detections with both $^{13}$CO ($J$=1--0) and C$^{18}$O ($J$=1--0) more than 5$\sigma$, we still can see that the maximum of $X_{\rm ^{13}CO}/X_{\rm C^{18}O}$ value occurs between $A_{\rm V} \sim$ 1--4\,mag.
To figure out the averaged trend of the $X_{\rm ^{13}CO}/X_{\rm C^{18}O}$ value, we averaged $X_{\rm ^{13}CO}/X_{\rm C^{18}O}$ value in 2.5\,mag $A_{\rm V}$ bins.
The averaged $X_{\rm ^{13}CO}/X_{\rm C^{18}O}$ value reaches the maximum of 10.3 which is $\sim$2 times more than 5.5 in the $A_{\rm V}$ range less than 2.5\,mag. 
Theoretical studies predict that in the mean ISRF, the $X_{\rm ^{13}CO}/X_{\rm C^{18}O}$ maximum occurs at $A_{\rm V} \sim$ 1--3\,mag and reaches $\sim$5--10 times of the intrinsic abundance ratio, and then decreases toward the intrinsic abundance ratio when $A_{\rm V}$ further increases \citep{war96}; 
that is, the selective photodissociation of C$^{18}$O appears at $A_{\rm V} \sim$ 1--3\,mag.
Our maximum of the averaged $X_{\rm ^{13}CO}/X_{\rm C^{18}O}$ value is only $\sim$2 times of 5.5, and our maximum of the non-averaged $X_{\rm ^{13}CO}/X_{\rm C^{18}O}$ value just barely reaches 27.7 which is $\sim$5 times of 5.5.
Since the model calculations in \citet{war96} only considered non-turbulent uniform density model, our lower maximum $X_{\rm ^{13}CO}/X_{\rm C^{18}O}$ value may be caused by non-uniform density distributions. 
Another example is shown by \citet{lad94} who derived a similar correlation in a filament of IC 5146 \citep[see Fig.\,19 in ][]{lad94}, which is also a low mass star-forming region. They found that the $X_{\rm ^{13}CO}/X_{\rm C^{18}O}$ ratio at $A_{\rm V}<$ 10\,mag is $\sim$3 times of 5.5 and then decreases to $\sim$5.5 at $A_{\rm V}>$ 10\,mag.
Note that the $X_{\rm ^{13}CO}/X_{\rm C^{18}O}$ peak positions in the $A_{\rm V}$ axes in L\,1551 and IC 5146 are different.
Although \citet{lad94} used the Near-Infrared Color Excess (NICE) technique to derive $A_{\rm V}$ and we used the gray-body SED fitting and Equ.\,\ref{equ_Av} to derive $A_{\rm V}$, both the NICE method and our method are based on the assumption that $R_{\rm V} = A_{\rm V}/E(B-V) = 3.1$.
Thus these difference may be owing to that the difference beam size between our data or there are still some different physical condition between them.

As seen in Fig.\,\ref{fig_intro} (c), the C$^{18}$O emission is rather weak or not detected at the periphery of the cloud, in which higher $X_{\rm ^{13}CO}/X_{\rm C^{18}O}$ value is expected. 
On the other hand, the $^{13}$CO emission is strong at these regions.
Since we only calculated $X_{\rm ^{13}CO}/X_{\rm C^{18}O}$ value as the signal-to-noise ratio of both $^{13}$CO and C$^{18}$O greater than 5 (see \S \ref{result_col_abu}), we may only sample the positions with high C$^{18}$O abundance.
In order to check this possibility, we can estimate the upper limit of $X_{\rm ^{13}CO}/X_{\rm C^{18}O}$ value by using the C$^{18}$O emission with the signal-to-noise ratio of 3 and assuming FWHM$_{\rm C^{18}O}$ as the value at the outskirts in Fig.\,\ref{fig_intro} (c) ($\sim$0.18\,km s$^{-1}$) at the region where the C$^{18}$O emission is weak ($\lesssim$ 5$\sigma$ level noise) but the $^{13}$CO emission is still strong ($\gtrsim$ 5$\sigma$ level noise).
Then $X_{\rm ^{13}CO}/X_{\rm C^{18}O}$ value can reach $\sim$45 at these regions, which is $\sim$8 times of 5.5.
	
\subsection{Influence of the excitation temperature}\label{temperature}

\begin{figure}
	\includegraphics[scale=.6]{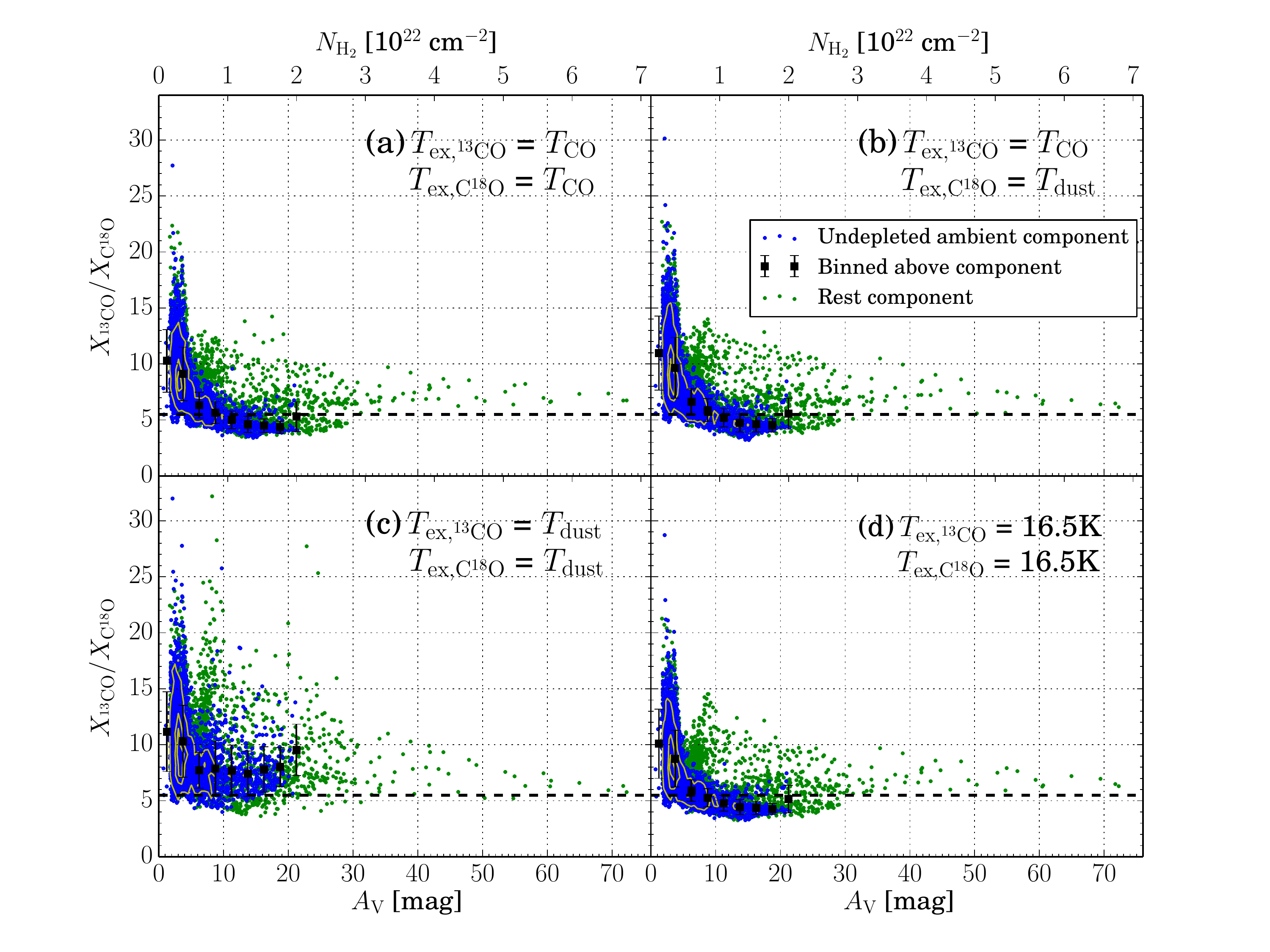}
	\caption{Correlation between the visual extinction, $A_{\rm V}$, and the abundance ratio, $X_{\rm ^{13}CO}/X_{\rm C^{18}O}$. The top x-axis is the column density of H$_2$, $N_{\rm H_2}$. The four panels (a), (b), (c), and (d) correspond to the four cases of the excitation temperatures of the $^{13}$CO and C$^{18}$O ($J$=1--0) lines. The blue and green dots denote the undepleted ambient and rest (depletion factor $>$ 1 and outflow regions) components, respectively. The black filled squares and their error-bars are the average values and the dispersion of $X_{\rm ^{13}CO}/X_{\rm C^{18}O}$ in the 2.5\,mag $A_{\rm V}$ bin. The yellow contours indicate the surface density of ambient points at the 15$\%$, 45$\%$, and 75$\%$ levels of the maximum surface density of the undepleted ambient points. The horizontal dashed lines indicate the solar system value of 5.5.}\label{fig_Av_XX}
\end{figure}

In order to derive the optical depths and column densities, we assumed that the $^{12}$CO, $^{13}$CO, and C$^{18}$O lines trace the same region, and derived the excitation temperature, $T_{\rm CO}$, using the peak intensity of $^{12}$CO ($J$=1--0). 
However, the three lines may trace different regions, and $T_{\rm CO}$ could not be used as the excitation temperatures of $^{13}$CO ($J$=1--0) and C$^{18}$O ($J$=1--0).
To investigate the influence of the excitation temperature, we re-estimate the abundance ratio by using the dust temperature (shown in Fig.\,\ref{fig_intro} (f)), $T_{\rm dust}$, as the excitation temperature of $^{13}$CO or/and C$^{18}$O. 
Since the thermal emission from dust is usually optically thin and can traces inner regions, the $^{13}$CO or/and C$^{18}$O may trace the same region as dust. 
We found that $T_{\rm dust}$ is generally smaller than $T_{\rm CO}$ in the inner region of L\,1551 
and $T_{\rm dust}$ in the area we calculated abundance ratio has a mean value of 13.2$\pm$0.7\,K; 
thus, there is a possibility that the $^{12}$CO emission and the dust emission may trace different regions in the inner region of L\,1551. 
The H$_2$ column density traced by the dust emission is also a very useful measure to examine whether the CO lines trace the same region as the dust or not. 
In fact, we can see the distributions of the $N_{\rm H_2}$ map and the $N_{\rm C^{18}O}$ map are similar, compared to the $N_{\rm ^{13}CO}$ map. 
Thus, we assume the excitation temperature of C$^{18}$O is the dust temperature, but we try both $T_{\rm CO}$ and $T_{\rm dust}$ for $^{13}$CO for comparison.
On the other hand, since \citet{sto06} used the two rotational transitions of $^{12}$CO ($J$=1--0 and 3--2) to derive a mean excitation temperature of the main NE/IRS 5 outflow to be 16.5\,K, 
which is generally higher than $T_{\rm CO}$, we also use this temperature for comparison.
Therefore, we compare the derived abundance ratios under the following four assumptions:
\begin{enumerate}
\item $T_{\rm ex, ^{13}CO}$ = $T_{\rm CO}$ and $T_{\rm ex, C^{18}O}$ = $T_{\rm CO}$. (Original)\label{Tex_1}
\item $T_{\rm ex, ^{13}CO}$ = $T_{\rm CO}$ and $T_{\rm ex, C^{18}O}$ = $T_{\rm dust}$.\label{Tex_2}
\item $T_{\rm ex, ^{13}CO}$ = $T_{\rm dust}$ and $T_{\rm ex, C^{18}O}$ = $T_{\rm dust}$.\label{Tex_3}
\item $T_{\rm ex, ^{13}CO} = T_{\rm ex, C^{18}O} = 16.5\,\rm K$.\label{Tex_4}
\end{enumerate}
Each panel in Fig.\,\ref{fig_Av_XX} shows the correlation between $X_{\rm ^{13}CO}/X_{\rm C^{18}O}$ and $A_{\rm V}$ under one of the above four assumptions. 
The mean and standard deviation of the abundance ratios are 8.0$\pm$2.8, 8.4$\pm$3.1, 9.5$\pm$3.2, and 7.6$\pm$2.9 under these four assumptions respectively.

The dependence of the $X_{\rm ^{13}CO}/X_{\rm C^{18}O}$ value on the $A_{\rm V}$ value under the assumption (\ref{Tex_2}) and (\ref{Tex_4}) are consistent with that under the original assumption (\ref{Tex_1}).
We notice that the $X_{\rm ^{13}CO}/X_{\rm C^{18}O}$ value at $A_{\rm V} \gtrsim$ 5 under the assumption (\ref{Tex_4}) is sightly lower than that in the assumption (\ref{Tex_1}), 
which reveals that a higher excitation temperature will derive lower abundance ratio.
Actually, $T_{\rm CO}$ used in Fig.\,\ref{fig_Av_XX} (a) has a mean value of 15.3$\pm$1.9\,K. 
Even if we use 16.5\,K as the excitation temperature in the assumption (\ref{Tex_4}), we still obtain almost the same trend in the correlation between $X_{\rm ^{13}CO}/X_{\rm C^{18}O}$ and $A_{\rm V}$.
Thus, this derivation of abundance ratio is not sensitive to the excitation temperature.
In the case of the assumption (\ref{Tex_3}), the trend of $X_{\rm ^{13}CO}/X_{\rm C^{18}O}$ is similar to those under the other assumptions in the $A_{\rm V}$ range less than 10\,mag, but the $X_{\rm ^{13}CO}/X_{\rm C^{18}O}$ value increases in the $A_{\rm V}$ range more than 10\,mag.
In fact, the ambient region in which the $A_{\rm V}$ value larger than 10\,mag is corresponding to L\,1551 MC in which the dust temperature is less than $\sim$10\,K.
Since the only difference between the assumption (\ref{Tex_2}) and (\ref{Tex_3}) is the excitation temperature of $^{13}$CO, 
this behavior can be interpreted as follows: the excitation temperature of $^{13}$CO in L\,1551 MC is underestimated in the case of $T_{\rm ex}$ = $T_{\rm dust}$, and thus the column density of $^{13}$CO is overestimated.
This suggests that the $^{13}$CO line does not trace the dense region of L\,1551 MC traced by the dust emission: L\,1551 MC cannot be recognized in the $^{13}$CO integrated intensity map of Fig. \ref{fig_intro} (b).

Consequently, even when we take into account the uncertainties in the excitation temperature estimation, our results still lead to the same conclusion.
That is, the $X_{\rm ^{13}CO}/X_{\rm C^{18}O}$ value reaches $\sim$11 at the low $A_{\rm V}$ value ($A_{\rm V} \sim$ 1--4\,mag) which is indeed larger than the solar system value, and decreases toward the solar system value in the high $A_{\rm V}$ value ($A_{\rm V} \sim$ 5--20\,mag) except the case of the low ($\sim$10 K) excitation temperature for the $^{13}$CO line.

\subsection{Comparison between the L\,1551 molecular cloud and the Orion-A giant molecular cloud}

Orion-A GMC is located at a distance of 400\,pc \citep{men07, san07, hir08} and is under strong influence of the FUV emission from the Trapezium cluster and NU Ori \citep{shi11}. 
Several photon dissociation regions (PDRs) have been identified from comparison of the distributions of the 8\,$\mu$m, 1.1\,mm, and $^{12}$CO ($J$=1--0) line emission \citep{hol97, shi11, shi13}. 
\citet{shi14} found that the abundance ratios of $^{13}$CO to C$^{18}$O in PDRs and non-PDRs are both higher than 5.5, and concluded that these high abundance ratios are due to the selective FUV photodissociation of C$^{18}$O. 
	
\begin{figure}
	\includegraphics[scale=.85]{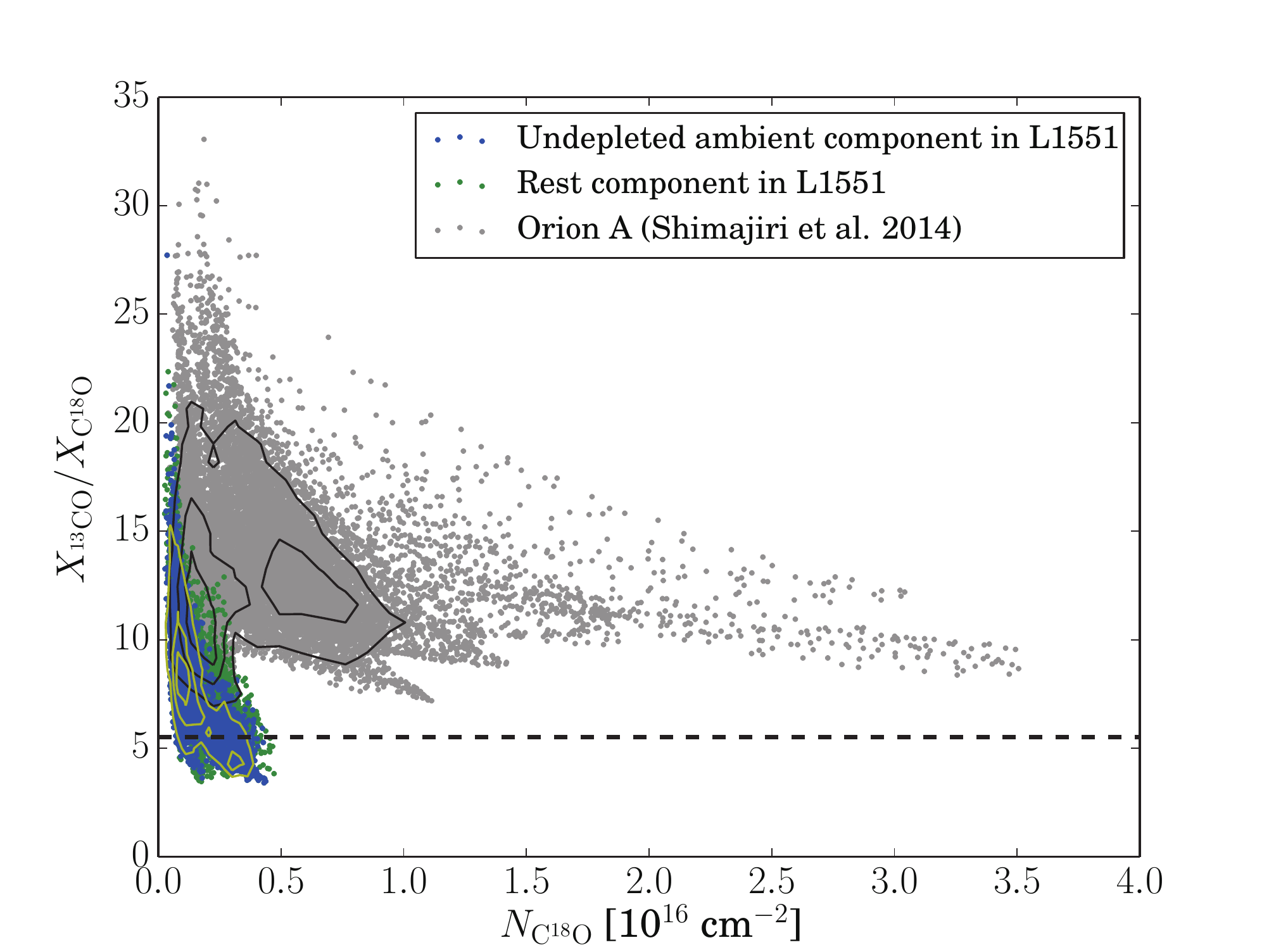}
	\caption{Correlation between the column density of C$^{18}$O, $N_{\rm C^{18}O}$, and the abundance ratio, $X_{\rm ^{13}CO}/X_{\rm C^{18}O}$ in L\,1551 and Orion A. The blue and green dots denote the undepleted ambient and rest components in L\,1551, respectively. The gray dots denote the Orion A data from Shimajiri et al. (2014).  The yellow and black contours indicate the surface density of the blue and grey points at the 15$\%$, 45$\%$, and 75$\%$ levels of their maximum surface density of the ambient component in L 1551 and Orion A, respectively. The horizontal dashed line indicates the solar system value of 5.5.}
	\label{fig_N18_XX}
\end{figure}

In order to investigate the influence of the different environments on the abundance ratio of $^{13}$CO to C$^{18}$O, we compare the abundance ratio between L\,1551 and Orion-A. 
Figure \ref{fig_N18_XX} shows the $X_{\rm ^{13}CO}/X_{\rm C^{18}O}$ value as a function of $N_{\rm C^{18}O}$ for L\,1551 and Orion-A (both the PDR and non-PDR regions).
The data of Orion-A are from \citet{shi14}.
Here, \citet{shi14} used the C$^{18}$O column density to estimate the total column density.
Although the C$^{18}$O column density and the $A_{\rm V}$ value do not have a perfect linear relation because of the selective photodissociation \citep{fre82}, 
we can still consider that the higher column density is generally corresponding to the higher $A_{\rm V}$ value and vice versa.
Our results show that the abundance ratio in L\,1551 is generally lower than that in Orion-A. 
The mean $X_{\rm ^{13}CO}/X_{\rm C^{18}O}$ value in L\,1551 is 8.0$\pm$2.8 in contrast to 16.5$\pm$0.07 in PDRs and 12.3$\pm$0.02 in non-PDRs in Orion-A \citep{shi14}.
For the non-PDRs of Orion-A, the $X_{\rm ^{13}CO}/X_{\rm C^{18}O}$ value has a maximum in the low $N_{\rm C^{18}O}$ regime ($<$ $5\times10^{15}$ cm$^{-2}$), and then decreases to $\sim$10 in the high $N_{\rm C^{18}O}$ regime \citep[see Fig.\,7 (d), (e), (f), and (g) in][]{shi14}. 
This trend is similar to the result of L\,1551
but the selective photodissociation of C$^{18}$O in L\,1551 only occurs at the outskirts ($A_{\rm V} \sim$ 1--4\,mag, see Fig.\,\ref{fig_Av_XX} (a)). 
However, even at the high column density, the abundance ratio is still about two times higher than 5.5 in Orion-A, because of the presence of the embedded OB stars in the cloud.
In the other words, the FUV strength in Orion-A is higher than that in L\,1551,
 which is the environmental difference between L\,1551 and Orion-A. 
In the Orion-A, the FUV radiation from the embedded OB stars is likely to penetrate the whole cloud.  
For the PDRs of Orion-A, the stronger FUV radiation causes the even higher convergence value of $\sim$15.

\subsection{CO-to-H$_2$ conversion factor across the L\,1551 molecular cloud}

Measurements of the mass distribution in molecular clouds help to understand their physical and chemical characteristics. However, in ISM, the most abundant molecular species, H$_2$, is hard to be observed because H$_2$ lacks its electric-dipole moment and its quadruple transition is hard to occur in the typical molecular cloud environment. The secondary abundant molecular species, CO, is not the case, because CO is much easier to be observed and the $^{12}$CO ($J$=1--0) emission is considered as the most available mass tracer. Thus, the CO-to-H$_2$ conversion factor, also called ``X-factor", is defined as,
\begin{equation}
{\rm X\textendash factor}=\frac{N_{\rm H_2}}{W_{\rm ^{12}CO}} \,{\rm cm^{-2}\,K^{-1}\,km^{-1}\,s},
\end{equation}
where $W_{\rm ^{12}CO}$ is the integrated intensity of the $^{12}$CO ($J$=1--0) emission. 
\citet{bol13} showed that the averaged X-factor = 2$\times$10$^{20}$\,cm$^{-2}$\,K$^{-1}$\,km$^{-1}$\,s with $\pm$30$\%$ uncertainty in the Milky Way disk.
Nevertheless, the X-factor can vary by a factor of $\sim$ 100 in different regions, 
because $^{12}$CO ($J$=1--0) is often optically thick \citep{lee14}. 
For example, 
\citet{pin10} measured X-factor $\sim$ (1.6--12)$\times$10$^{20}$\,cm$^{-2}$\,K$^{-1}$\,km$^{-1}$\,s in the Taurus molecular cloud, 
\citet{lee14} measured X-factor $\sim$ 3$\times$10$^{19}$\,cm$^{-2}$\,K$^{-1}$\,km$^{-1}$\,s in the Perseus molecular cloud, 
and \citet{kon15} measured X-factor $\sim$ 2.53$\times$10$^{20}$\,cm$^{-2}$\,K$^{-1}$\,km$^{-1}$\,s in the southeastern part of the California molecular cloud.
These discrepancies are also found in ISM numerical simulations which show that X-factor is likely dependent on extinction, volume density, temperature, metallicity, turbulence, star formation feedback, and so on \citep{she11a,she11b,lee14,cla15}. 
Even within a molecular cloud, the X-factor can still vary with a wide range.
On the other hand, the variation of the X-factor of the $^{13}$CO and C$^{18}$O can be smaller (see Appendix \ref{appendix_X}). 

\begin{figure}
	\includegraphics[scale=.6]{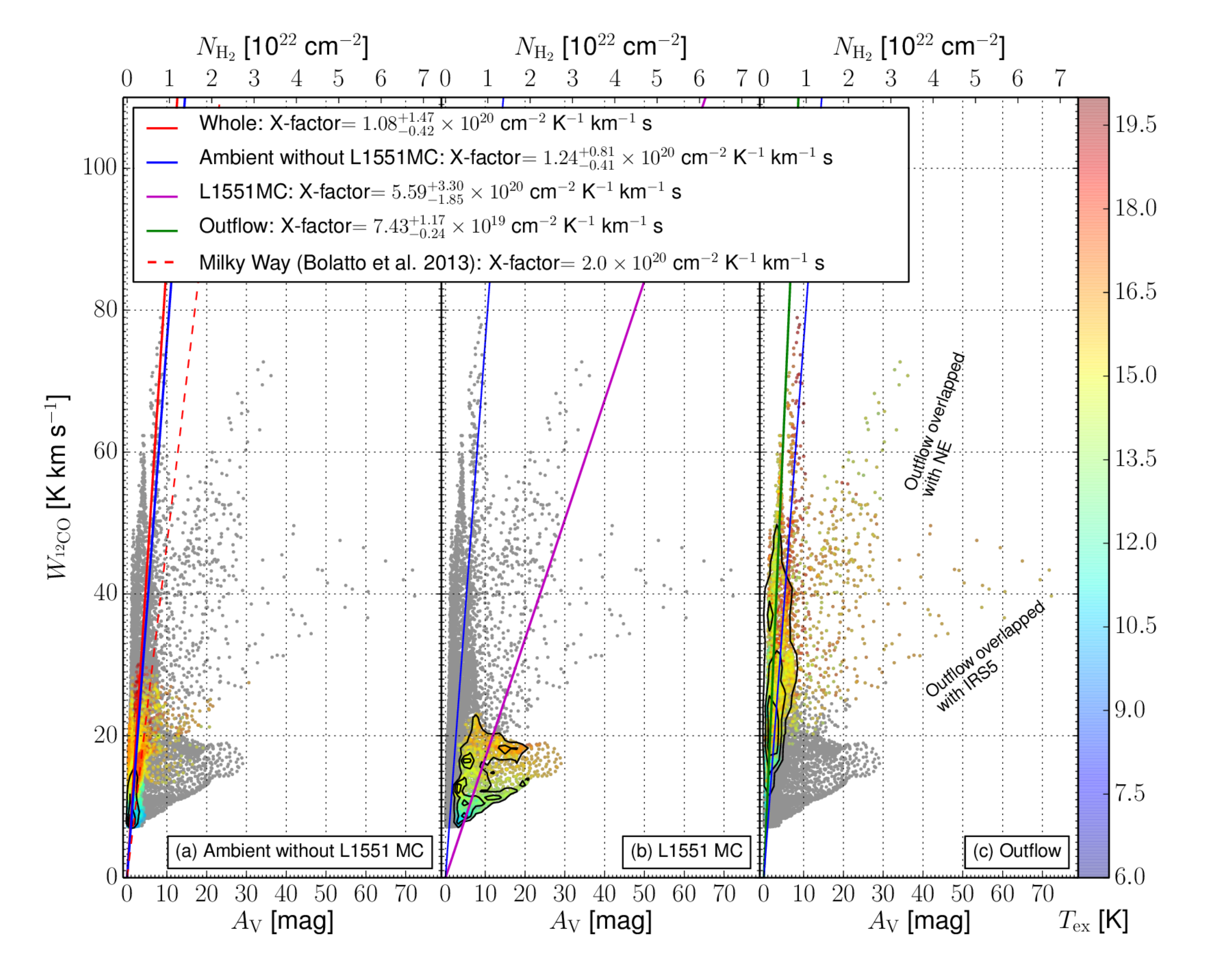}
	\caption{Correlation between the visual extinction, $A_{\rm V}$, and the integrated intensity of $^{12}$CO ($J$=1--0), $W_{\rm ^{12}CO}$. The gray dots denotes all data points in the map. The color-coded dots show the data points in the labeled regions, and the color represents the excitation temperature, $T_{\rm ex}$. The black contours indicate the surface density of the color-coded points at the 15$\%$, 45$\%$, and 75$\%$ levels of the maximum surface density in each panel.}
	\label{fig_Av_W12}
\end{figure}

Hereafter we calculate the X-factor for pixels in Fig.\,\ref{fig_intro} (a) with the signal-to-noise ratio of the $^{12}$CO integrated intensity greater than 3.
Figure \ref{fig_Av_W12} shows the correlation between the $W_{\rm ^{12}CO}$ value and the $A_{\rm V}$ value, and the points are color-coded by $T_{\rm ex}=T_{\rm CO}$. 
We can see that there are multiple trends. 
The averaged X-factor of the whole region is 1.08$^{+1.47}_{-0.42}\times$10$^{20}$\,cm$^{-2}$\,K$^{-1}$\,km$^{-1}$\,s. This is about a factor of two smaller than the average X-factor in Milky Way, but the dispersion still covers the average X-factor in Milky Way. 
In order to reveal these different trends, although we have already defined two regions, the ambient and outflow components, in the previous discussion, we divide into more regions based on the previous division here: 
(a) the diffuse component which is the ambient component excluding the L\,1551 MC component, 
(b) the L\,1551 MC component which is the green polygon in Fig.\,\ref{fig_XX} (f), 
and (c) the outflow component which is the same as defined in Sec. 4.1.
In each panel, we use the chi-squre fitting method with errors in both coordinates to find the best fit line.

The diffuse component (Fig.\,\ref{fig_Av_W12} (a)) shows a relatively narrow distribution, compared with the other two regions, and the fitted X-factor = 1.24$^{+0.81}_{-0.41}\times$10$^{20}$\,cm$^{-2}$\,K$^{-1}$\,km$^{-1}$\,s is similar to the fitted X-factor of the whole region and is the same order of magnitude as the average X-factor in Milky Way.
The L\,1551 MC component (Fig.\,\ref{fig_Av_W12} (b)), however, shows a more extended distribution and a larger fitted X-factor = 5.59$^{+3.30}_{-1.85}\times$10$^{20}$\,cm$^{-2}$\,K$^{-1}$\,km$^{-1}$\,s.
This is due to the fact that the $^{12}$CO emission is optically thick and thus saturated in this dense starless core. 
The outflow component (Fig.\,\ref{fig_Av_W12} (c)) has a fitted X-factor = 7.43$^{+1.17}_{-0.24}\times$10$^{19}$\,cm$^{-2}$\,K$^{-1}$\,km$^{-1}$\,s, which is smaller than that of the diffuse component, 
but shows three different distributions.
Two of the distributions are extended and have shallow slopes in the correlation, which belong to the outflow region overlapped with IRS 5 and NE shown in Fig.\,\ref{fig_XX} (f) as two green incomplete circles and  marked in the panel (c) of Fig.\,\ref{fig_Av_W12}.
The $^{12}$CO emission from the IRS 5 and NE region is likely saturated because the H$_2$ column densities of IRS 5 and NE are comparable to that of L\,1551 MC ($N_{\rm H_2}\sim10^{22}$\,cm$^{-2}$). 
The other narrow distribution belongs to the remaining outflows and dominates the fitting of X-factor, which may be due to the fact that the entrained energy more strongly excites the $^{12}$CO emission and makes its high excitation temperature. 
Consequently, in the low $A_{\rm V}$ range ($A_{\rm V}\lesssim10$\,mag), the X-factor of the diffuse component is consistent with the Milky Way average value, and the X-factor of the outflow component is smaller. 
However, in L\,1551 MC ($A_{\rm V}\gtrsim3$\,mag), the X-factor becomes larger and has a large dispersion. 

\begin{figure}
	\includegraphics[scale=.6]{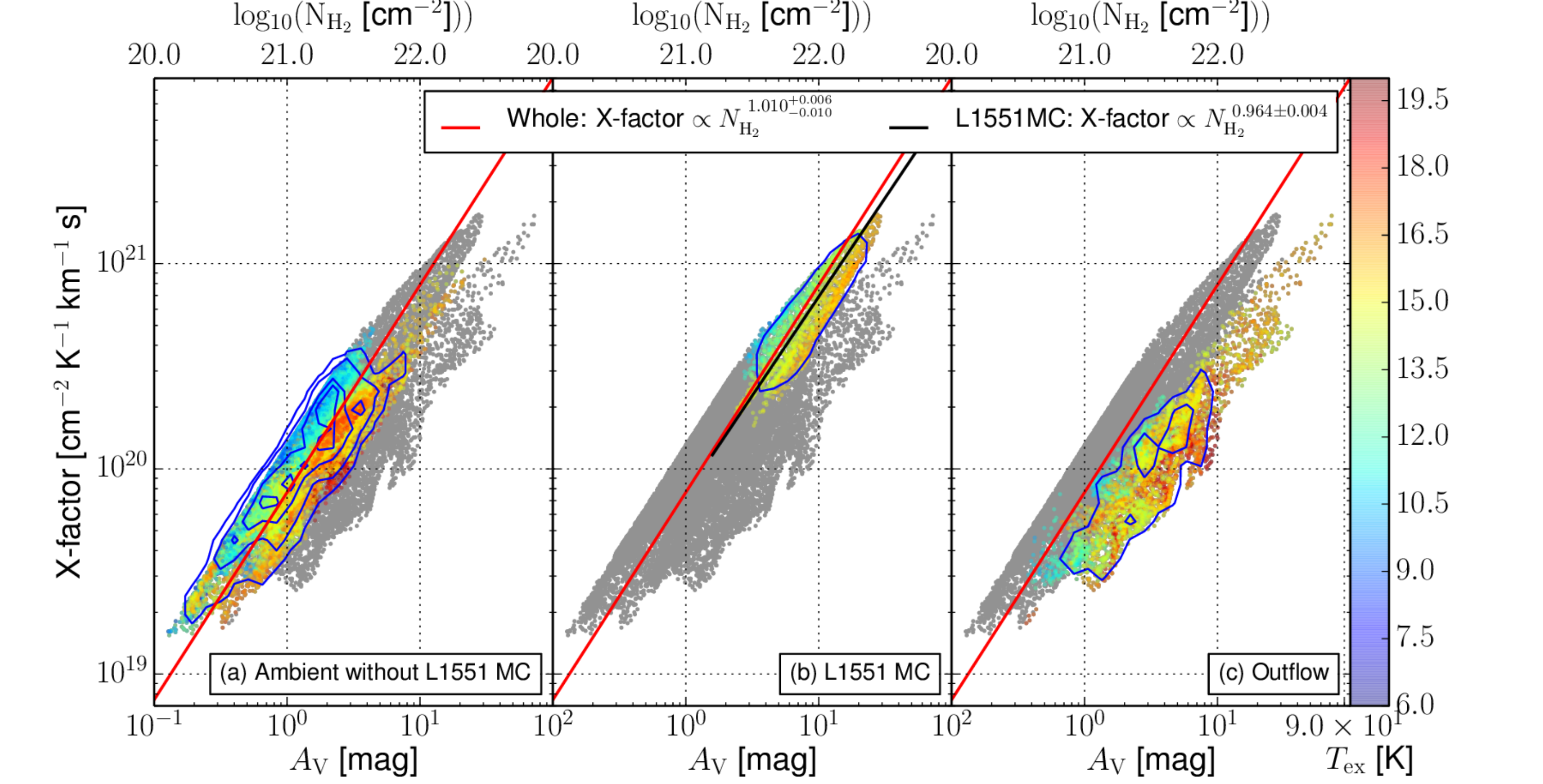}
	\caption{Correlation between the visual extinction, $A_{\rm V}$, and the X-factor. The gray dots denotes all data points in the map. The color-coded dots show the data points in the labeled regions, and the color represents the excitation temperature, $T_{\rm ex}$. The blue contours indicate the surface density of the color-coded points at the 5$\%$, 15$\%$, 45$\%$, and 75$\%$ levels of the maximum surface density of all data.}
	\label{fig_Av_X}
\end{figure}

Figure \ref{fig_Av_X} shows the correlation between the X-factor and the $A_{\rm V}$ value, and the points are color-coded by $T_{\rm ex}$. 
We can see that the X-factor of the whole region is dependent on $A_{\rm V}$ with a fitted power law, X-factor $\propto$ $N_{\rm H_2}^{1.01\pm0.01}$.
Especially, the L\,1551 MC component has a well-correlated distribution (less dispersion of the power index), X-factor $\propto$ $N_{\rm H_2}^{0.964\pm0.004}$.
A similar distribution but a shallower power law, X-factor $\propto$ $N_{\rm H_2}^{0.7}$, was also found by \citet{kon15} in the southeastern part of the California molecular cloud excluding the hot region around the massive star LkH$\alpha$ 101. 
However, in Perseus, \citet{lee14} found two characteristic features: the X-factor steeply decreases at $A_{\rm V}\lesssim3$\,mag and gradually increase at $A_{\rm V}\gtrsim3$\,mag. 
The steeply decreasing trend is likely due to the sharp transition from CII/CI to CO, but this steeply decreasing trend does not appear in our results. 
The ISM numerical simulations performed by \citet{she11a} suggests that the transition behavior is dependent on the volume density. 
They simulated molecular clouds with uniform density distribution from 10$^2$ to 10$^3$\,cm$^{-3}$ and initial turbulence, 
and showed that the low density case has the decreasing trend with a wider $A_{\rm V}$ range 
compared to the high density case.
For the highest density case in their simulation (n $\sim$ 10$^3$\,cm$^{-3}$), the result only shows one characteristic trend, 
X-factor $\propto$ $N_{\rm H_2}$, which means the $^{12}$CO emission is saturated (optically thick) to be constant and then the X-factor directly relates to $N_{\rm H_2}$, which is consistent with our results.
In L\,1551, we do not find the decreasing trend even at the low $A_{\rm V}$ range, which indicates the FUV strength in L\,1551 is less than that in Perseus. Under the mean ISRF, $^{12}$CO is self-shielded at $A_{\rm V}\sim$ 0.5\,mag \citep{rol13,szu14}, which may be the case in L\,1551.

\begin{figure}
	\includegraphics[scale=.6]{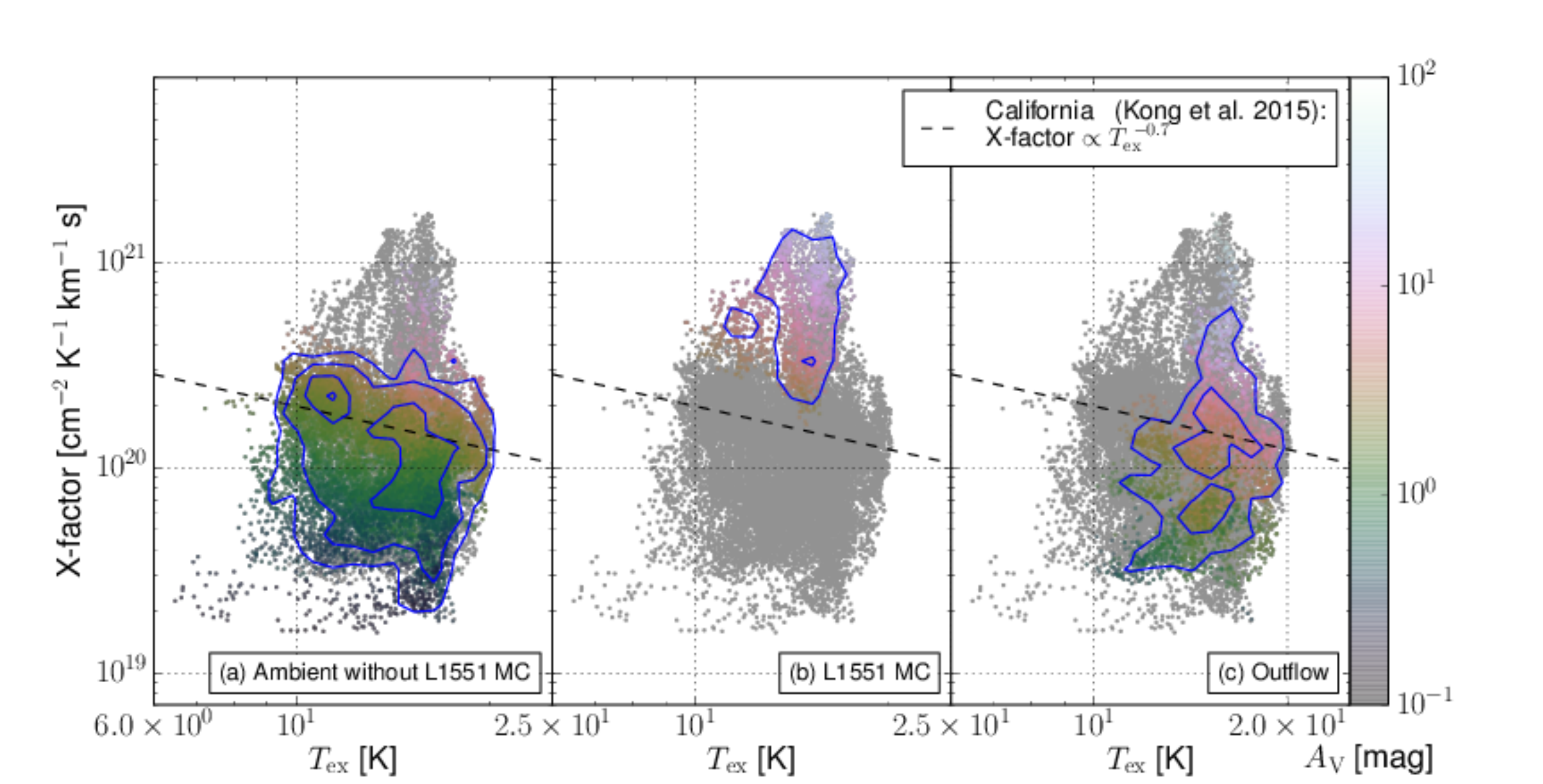}
	\caption{Correlation between the excitation temperature, $T_{\rm ex}$, and the X-factor. The gray dots denotes all data points in the map. The color-coded dots show the data points in the labeled regions, and the color represents the visual extinction, $A_{\rm V}$. The blue contours indicate the surface density of the color-coded points at the 5$\%$, 15$\%$, 45$\%$, and 75$\%$ levels of the maximum surface density of all data.}
	\label{fig_Tex_X}
\end{figure}

Figure \ref{fig_Tex_X} shows the correlation between the X-factor and the $T_{\rm ex}$ value, and the points are color-coded by $A_{\rm V}$.  
Our data do not show obvious correlated distributions.
However, an inverse power law, X-factor = $2\times10^{20}(T_{\rm ex}/10)^{-0.7}$, was found in the southeastern part of California cloud \citep{kon15}, 
which is consistent with X-factor $\propto$ $T_{\rm ex}^{-0.5}$ in simulations of \citet{she11b}.
Although this relation passes through the diffuse component of L\,1551, 
we can not confirm this relationship due to the narrow  temperature range of our data.

\begin{figure}
	\includegraphics[scale=.6]{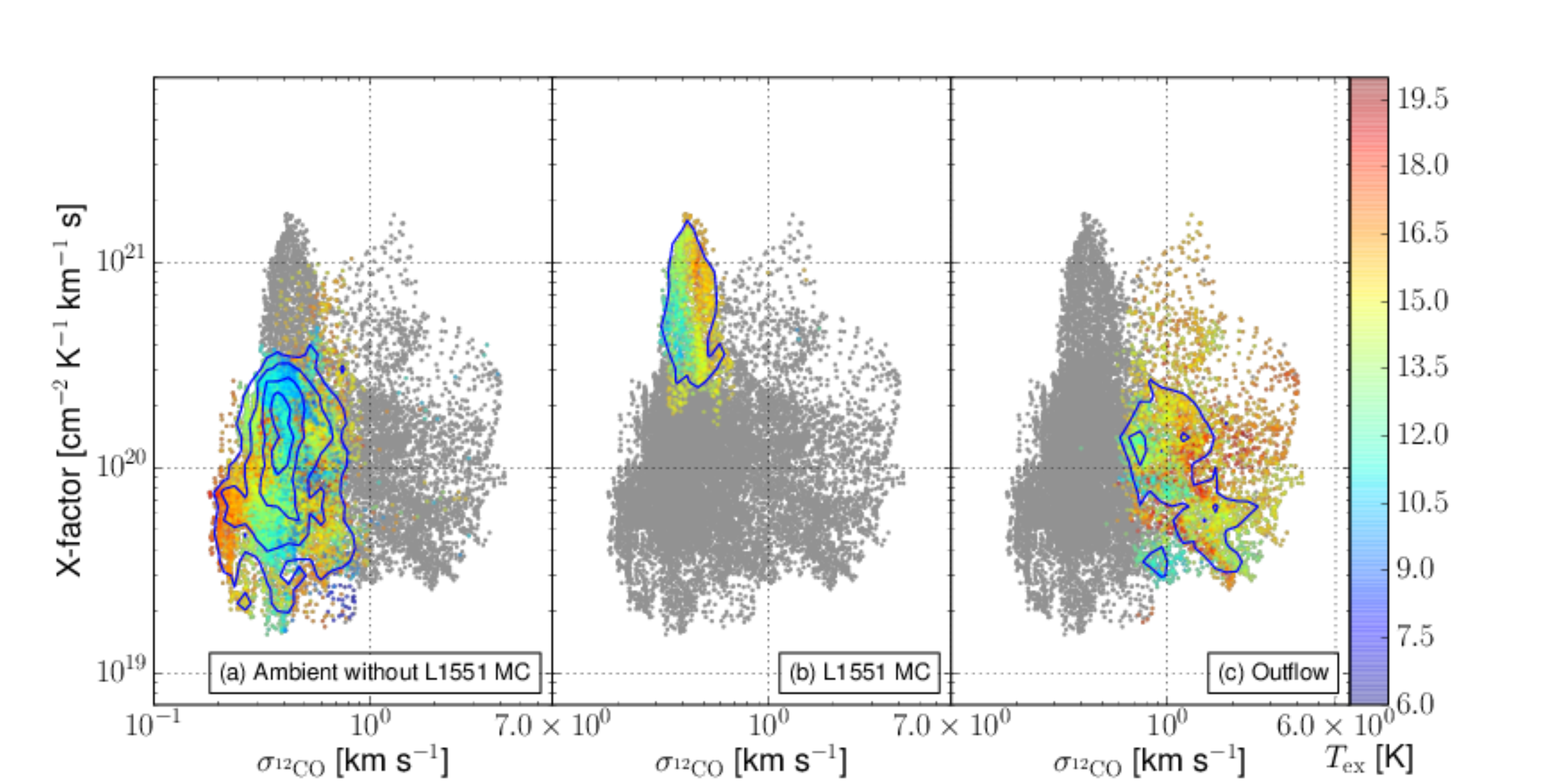}
	\caption{Correlation between the velocity dispersion of $^{12}$CO ($J$=1--0), $\sigma_{\rm ^{12}CO}$, and the X-factor. The gray dots denotes all data points in the map. The color-coded dots show the data points in the labeled regions, and the color represents the excitation temperature, $T_{\rm ex}$. The blue contours indicate the surface density of the color-coded points at the 5$\%$, 15$\%$, 45$\%$, and 75$\%$ levels of the maximum surface density of all data.}
	\label{fig_mom2_X}
\end{figure}

Figure \ref{fig_mom2_X} shows the correlation between the X-factor and the $^{12}$CO velocity dispersion, $\sigma_{\rm ^{12}CO}$, and the points are color-coded by $T_{\rm ex}$. 
The simulations predict that X-factor $\propto$ $\sigma_{\rm ^{12}CO}^{-0.5}$ \citep{she11b}.
However, 
our data do not have a correlated distribution.
\citet{kon15} also found no correlation in the southeastern part of California cloud.

\section{Summary}\label{sec_summary}

We have carried out wide-field OTF observations in the $^{12}$CO, $^{13}$CO, and C$^{18}$O ($J$=1--0) lines toward the L\,1551 molecular cloud using the BEARS receiver of the NRO 45\,m telescope. 
The main results are summarized as follows:

\begin{enumerate}

\item The abundance ratio of $^{13}$CO to C$^{18}$O ($X_{\rm ^{13}CO}/X_{\rm C^{18}O}$ = $N_{\rm ^{13}CO}/N_{\rm C^{18}O}$) is derived to be $\sim$3--27. 
The mean value is 8.0$\pm$2.8. 
We can see a spatially decreasing trend in the $X_{\rm ^{13}CO}/X_{\rm C^{18}O}$ map from the outskirts to the dense part of L\,1551, the starless core L\,1551 MC.

\item We found that the $X_{\rm ^{13}CO}/X_{\rm C^{18}O}$ value reaches its maximum in $A_{\rm V} \sim$ 1--4\,mag, and then decreases down to about the typical solar system value of 5.5 in the high $A_{\rm V}$ value, 
suggesting that the selective photodissociation of C$^{18}$O from the interstellar FUV radiation occurs in L\,1551.
PDR theoretical models with the mean ISFR suggest that the selective photodissociation of C$^{18}$O causes the increase of $X_{\rm ^{13}CO}/X_{\rm C^{18}O}$ value up to $\sim$5--10 times of 
the intrinsic value in the low $A_{\rm V}$ regions with $\sim$ 1--3\,mag.
However, in L\,1551, the average $X_{\rm ^{13}CO}/X_{\rm C^{18}O}$ value in 2.5\,mag bins reaches its maximum value of 10.3 which is $\sim$2 times of 5.5 in the $A_{\rm V}$ range less than 2.5\,mag.
Since the model calculations consider non-turbulent uniform cloud density, 
this lower maximum $X_{\rm ^{13}CO}/X_{\rm C^{18}O}$ value may be caused by non-uniform cloud density.
	
\item The mean of the excitation temperatures of $^{13}$CO and C$^{18}$O 
derived from $^{12}$CO in the non-outflow regions is 15.3$\pm$1.9\,K.
Even if we consider the higher excitation temperature of $T_{\rm ex, ^{13}CO}=T_{\rm ex, C^{18}O}=$ 16.5\,K or assume $T_{\rm ex, ^{18}CO}=T_{\rm dust}$, we also find a similar trends of $X_{\rm ^{13}CO}/X_{\rm C^{18}O}$. 
Therefore, our results described in item 2 are not sensitive to the uncertainties in the  excitation temperature determination.
	
\item The trends in the $X_{\rm ^{13}CO}/X_{\rm C^{18}O}$ variations in L\,1551 and Orion-A are similar to each other.
That is, both $X_{\rm ^{13}CO}/X_{\rm C^{18}O}$ values reach their maximums at low $N_{\rm C^{18}O}$ value, and then decrease as $N_{\rm C^{18}O}$ value increase; however, the $X_{\rm ^{13}CO}/X_{\rm C^{18}O}$ value in Orion-A converges to a higher value than that in L\,1551. 
This is due to the FUV radiation from the embedded OB stars.

\item We calculated that the averaged X-factor = $N_{\rm H_2}/W_{\rm ^{12}CO (\it J\rm=1-0)}$ in L\,1551 is 1.08$^{+1.47}_{-0.42}\times$10$^{20}$\,cm$^{-2}$\,K$^{-1}$\,km$^{-1}$\,s which is somewhat smaller but consistent with the Milky Way average value $\sim$ 2$\times$10$^{20}$\,cm$^{-2}$\,K$^{-1}$\,km$^{-1}$\,s.

\item The X-factor of the different regions in L\,1551 has a large variation. For the diffuse region, we found a X-factor = 1.24$^{+0.81}_{-0.41}\times$10$^{20}$\,cm$^{-2}$\,K$^{-1}$\,km$^{-1}$\,s which is similar to the Milky Way average value. 
For the outflow region, we found a smaller X-factor = 7.43$^{+1.17}_{-0.24}\times$10$^{19}$\,cm$^{-2}$\,K$^{-1}$\,km$^{-1}$\,s. The region of L\,1551 MC has a larger X-factor = 5.59$^{+3.30}_{-1.85}\times$10$^{20}$\,cm$^{-2}$\,K$^{-1}$\,km$^{-1}$\,s with a large dispersion because of the saturation of the $^{12}$CO emission. For the whole region, we found a correlation between the X-factor and the $A_{\rm V}$ value with X-factor $\propto$ $N_{\rm H_2}^{1.01\pm0.01}$ at $A_{\rm V}$ = $\sim$0.2--70\,mag which is due to the saturation of the $^{12}$CO emission. 

\end{enumerate}

\acknowledgments
The 45\,m radio telescope is operated by Nobeyama Radio Observatory, a branch of National Astronomical Observatory of Japan. 
Part of this work was achieved using the grant of NAOJ Visiting Fellow Program supported by the Research Coordination Committee, 
National Astronomical Observatory of Japan (NAOJ). 
S.J.L. and S.P.L. acknowledge support from the Ministry of Science and Technology of Taiwan with Grants MOST 102-2119-M-007-004-MY3. 

\appendix
\section{A. Velocity structure of the $^{12}$CO ($J$=1--0) emission line}\label{appendix_12CO}

Figure \ref{fig_ch_12CO} shows the velocity channel maps of $^{12}$CO ($J$=1--0). 
In the velocity range from $-$8\,km\,s$^{-1}$ to 6\,km\,s$^{-1}$, an extended elongated emission can be recognized at the southwest of IRS 5. This component is identified as the blueshifted lobe of the outflows ejected from IRS 5 and NE \citep{emr84, mor06, sto06}. 
In the velocity range from 4\,km\,s$^{-1}$ to 10\,km\,s$^{-1}$, a narrower elongated structure appears at the northeast of IRS 5 and through NE, which could still be a mix of the outflows from both sources. 
In the velocity range from 2\,km\,s$^{-1}$ to 6\,km\,s$^{-1}$, a clam-shaped emission shows up around the HL Tau group, which are identified as the collective outflows from the HL Tau group \citep{mun90, sto06, yos10}.  This outflow structure extends to $\sim$14\,km\,s$^{-1}$ in the redshifted lobe. 	
In the velocity range from 10\,km\,s$^{-1}$ to 20\,km\,s$^{-1}$, there is an emission between NE and the HL Tau group. 
This component is located between the redshifted outflows from IRS 5 and NE and the collective outflows from the HL Tau group \citep{sto06}, which could be the results of the outflow interaction.
In the velocity range from 8\,km\,s$^{-1}$ to 16\,km\,s$^{-1}$, another narrow elongated emission is distributed in the east--west direction pointing back to IRS 5 and NE, but the origin of this east--west outflow is still unclear \citep{mor91, pou91, rei02, sto06}. 
At 10\,km\,s$^{-1}$, a diffused emission with an intensity of $\sim$2--4\,K is distributed at the southeast of the L\,1551 molecular cloud, and this is a part of the ambient gas in Taurus \citep{yos10}. 
At 8\,km\,s$^{-1}$, two individual small cores with an extent of $\sim$3--5$\arcmin$ at the northern side probably do not associate with the L\,1551 molecular cloud because their velocity of $\sim$8\,km s$^{-1}$ is different from the systemic velocity of $\sim$6\,km s$^{-1}$ in L\,1551 (see the 6\,km s$^{-1}$ panel in Fig.\,\ref{fig_ch_12CO}).  

Figures \ref{fig_mom} (a) and (d) show $^{12}$CO mean velocity and velocity dispersion maps, respectively. The mean velocity over whole region is 6.6$\pm$0.44\,km\,s$^{-1}$ \citep{yos10}. We calculate that the mean velocity dispersion toward all observed area is 0.44$\pm$0.35\,km\,s$^{-1}$. 
Note that the velocity dispersion at the area of the outflows ($\sim$0.5--4.0\,km\,s$^{-1}$) is relatively higher than the mean value toward the overall observed area.

\section{B. Velocity structure of the $^{13}$CO ($J$=1--0) emission line}\label{appendix_13CO}

Figure \ref{fig_ch_13CO} shows the velocity channel maps of $^{13}$CO ($J$=1--0). In the velocity range from 4.25\,km\,s$^{-1}$ to 7.25\,km\,s$^{-1}$, we can see the U-shaped wall clearly at the southwest of IRS 5 
which is distributed around the southwestern blueshifted lobe of the outflows traced in $^{12}$CO and extending to the outside of those outflows.
Note that although the U-shaped wall seems to surround the blueshifted outflows, its velocity covers both redshifted and blueshifted ranges, which may hint that the blueshifted outflow axis is almost parallel to the plane of sky.
In the velocity range from 6.25\,km\,s$^{-1}$ to 7.25\,km\,s$^{-1}$, we can see the cavity at the northeast of NE. 
At the velocity of 6.25\,km\,s$^{-1}$, we can see a narrow filamentary structure sticking out from IRS 5 and extending toward L\,1551 MC and beyond. 
At the velocity of 6.75\,km\,s$^{-1}$, the $^{13}$CO emission is distributed over the whole region, which is consistent to the centroid velocity of 6.7$\pm$0.21\,km\,s$^{-1}$ measured by \citet{yos10}. 
At the velocity of 7.75\,km\,s$^{-1}$, we can barely see an east--west elongated structure corresponding to the east--west outflow seen in the $^{12}$CO map. 

Figure \ref{fig_mom} (b) shows the $^{13}$CO mean velocity map. 
The cavity is clearly seen in the blueshifted velocities with respect to the mean velocity of 6.7\,km s$^{-1}$.  Because the cavity is on top of the redshifted outflows from IRS 5/NE, we speculate that the gas on the cavity region is pushed toward us by the redshifted outflows. 
There is a redshifted streamer ($v_{\rm LSR}\gtrsim6.8\rm \,km\,s^{-1}$, see the black dashed line in Fig.\,\ref{fig_mom} (b)) at the southwest of the narrow filamentary structure (see the grey dashed line in Fig.\,\ref{fig_mom} (b)) and at the north of east--west outflow. However, the integrated intensity of this streamer is very low in Fig.\,\ref{fig_intro} (b), an thus this component is not discussed hereafter.
Figure \ref{fig_mom} (e) shows the $^{13}$CO velocity dispersion map.  
The velocity dispersion of the U-shaped wall ($\sim$0.4--0.7\,km\,s$^{-1}$) is relatively higher than the mean value in the overall observed area (0.20$\pm$0.09\,km\,s$^{-1}$).

\section{C. Velocity structure of the C$^{18}$O ($J$=1--0) emission line}\label{appendix_C18O}

Figure \ref{fig_ch_C18O} shows the velocity channel maps of C$^{18}$O ($J$=1--0). 
These maps show a filamentary structure in the northwest--southeast direction in the velocity range from 6.25\,km\,s$^{-1}$ to 7.25\,km, and L\,1551 MC traced by NH$_3$ coincides with this filament \citep{swi05}.	  	
This filamentary structure can also be recognized in the $^{13}$CO channel maps (the 6.25\,km\,s$^{-1}$ panel of Fig.\,\ref{fig_ch_13CO}) and the H$_2$ column density map (Fig.\,\ref{fig_intro} (d)). 
A U-shaped wall structure is present in the velocity range from 6.75\,km\,s$^{-1}$ to 7.05\,km, which coincides with the redshifted part of the U-shaped wall traced in $^{13}$CO. 
Another fainter filamentary structure can be seen in the velocity range from 7.05\,km\,s$^{-1}$ to 7.35\,km\,s$^{-1}$, and its position is just above the upper arm of the U-shaped wall.  
Figure \ref{fig_mom} (c) and (f) are the C$^{18}$O mean velocity and the velocity dispersion maps, respectively.
Within the U-shaped wall, the integrated intensity becomes low and the gas is mostly blueshifted (see Fig.\,\ref{fig_intro} (c)).

\section{D. Relationship between CO isotopes and H$_2$ column density}\label{appendix_X}

Figures \ref{fig_Av_W13} and \ref{fig_Av_W18}, respectively, show the correlation between the $W_{\rm ^{13}CO}$ value, the integrated intensity of the $^{13}$CO ($J$=1--0) emission, and the $A_{\rm V}$ value, and the $W_{\rm C^{18}O}$ value, the integrated intensity of the C$^{18}$O ($J$=1--0) emission, and the $A_{\rm V}$ value,  
in which the points are color-coded by $T_{\rm ex}=T_{\rm CO}$. 
In order to make direct comparison with their column density, 
we only plot the same regions where the pixels have their signal-to-noise ratio greater than 5 (see \S \ref{result_col_abu}). The fitted X-factor of the $^{13}$CO and C$^{18}$O emission are 7.48$^{+5.18}_{-2.64}\times$10$^{20}$ and 7.23$^{+3.41}_{-2.44}\times$10$^{21}$\,cm$^{-2}$\,K$^{-1}$\,km$^{-1}$\,s, respectively. 
In Fig.\,\ref{fig_Av_W13} and \ref{fig_Av_W18}, the distribution of points in each panel is similar to that for $^{12}$CO but has a narrower distribution because their optical depths are smaller than that of the $^{12}$CO emission.
Especially for the C$^{18}$O emission, since $\tau_{\rm C^{18}O}<0.8$, even the distribution of L\,1551 MC has a linear correlation, compared with the $^{12}$CO and $^{13}$CO.

\clearpage
\clearpage
\clearpage
\clearpage
\clearpage
\clearpage
\clearpage
\clearpage
\clearpage

\setcounter{figure}{0}
\renewcommand{\thefigure}{A\arabic{figure}}

\begin{figure}
	\includegraphics[scale=.7]{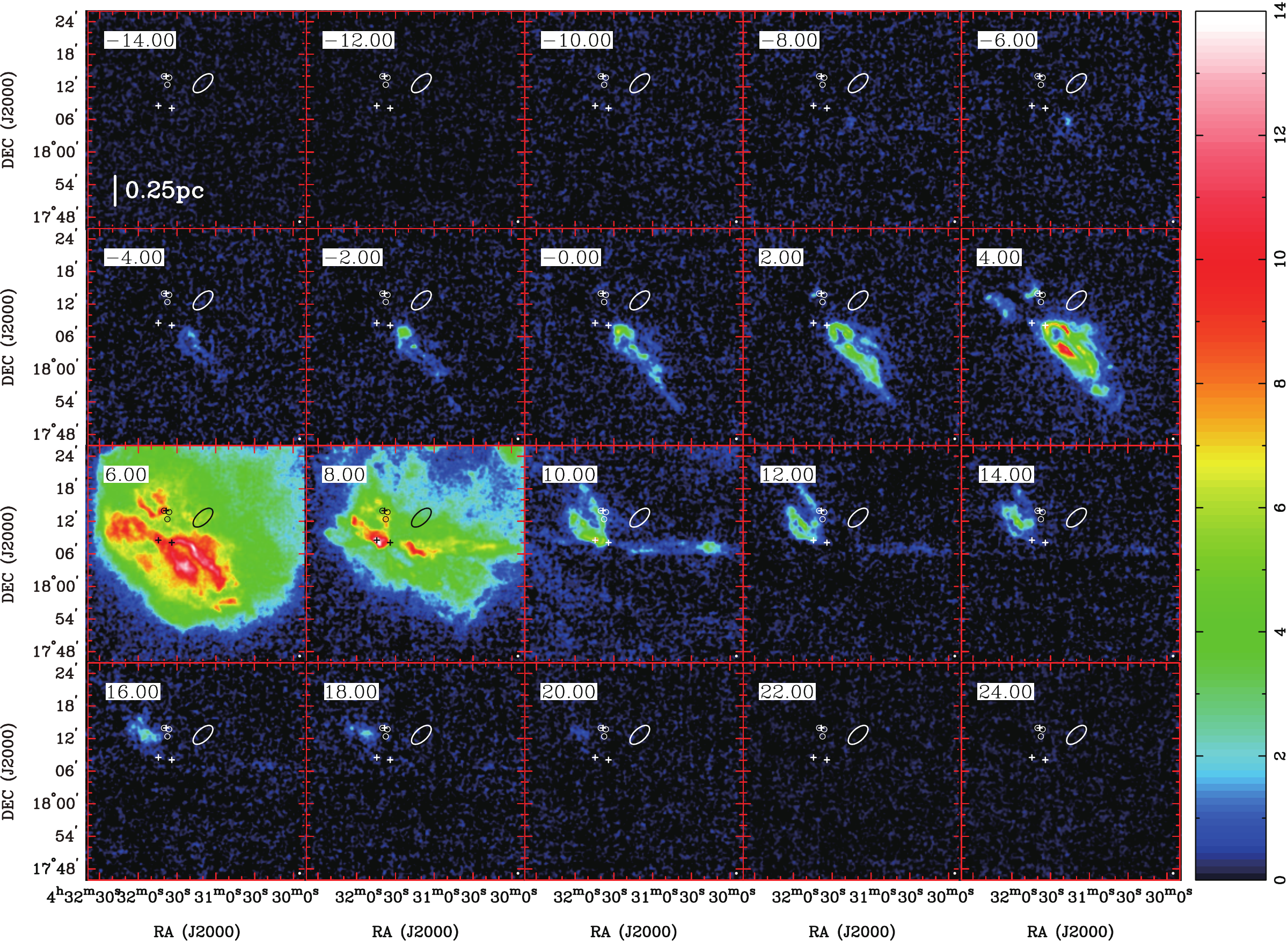}
	\caption{L\,1551 velocity channel maps of $^{12}$CO ($J$=1--0) in units of K. The numbers at the top-left corners denote the central  LSR velocities of the individual channels in units of km\,s$^{-1}$. The effective beam sizes are 30$\farcs$0. The linear scale of 0.25\,pc is shown at the bottom-left corner of the first panel.\label{fig_ch_12CO}}
\end{figure}
\clearpage

\begin{figure}
	\includegraphics[scale=.7]{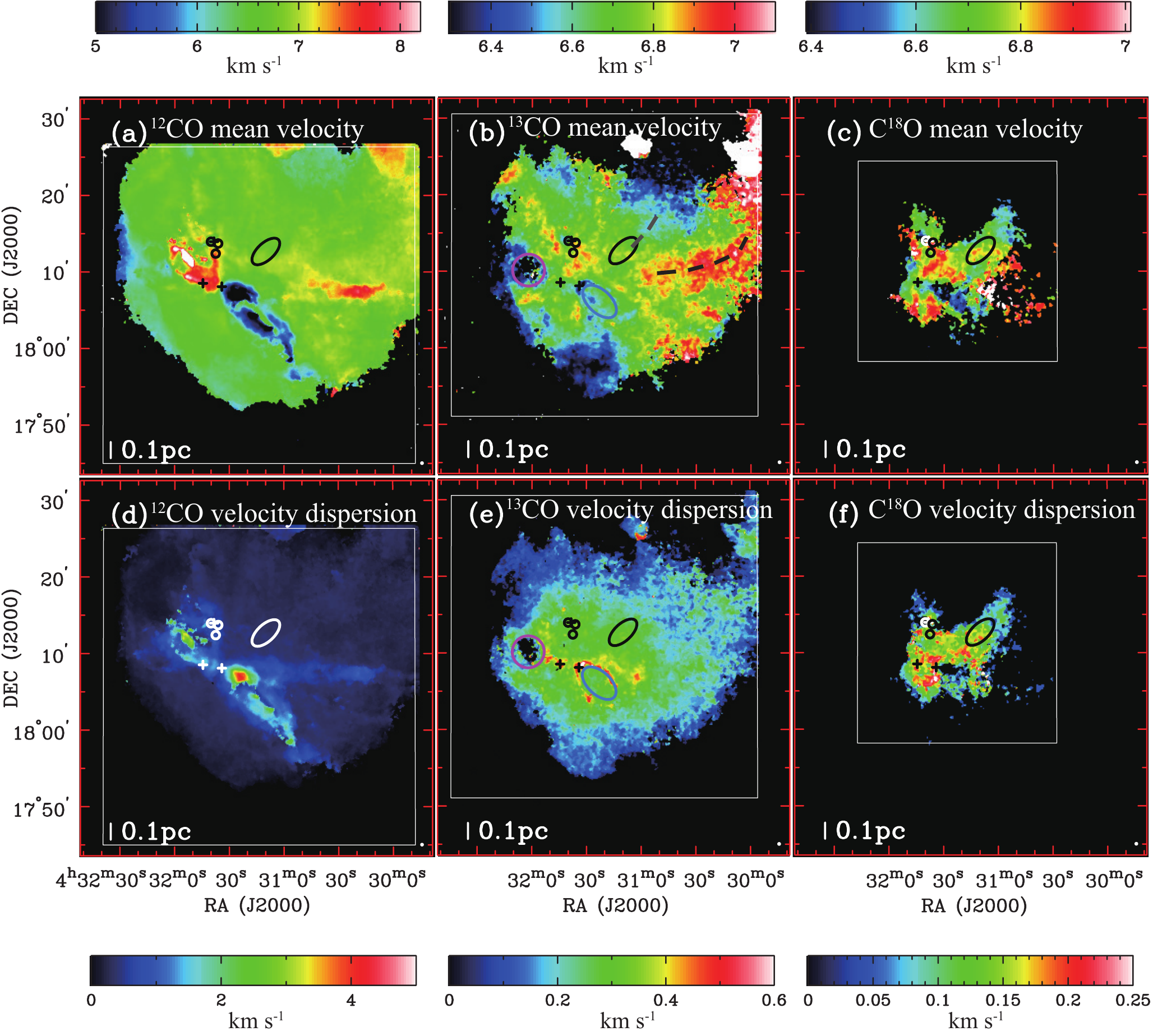}
	\caption{L\,1551 maps of mean velocity and velocity dispersion. The mean velocity maps of (a) $^{12}$CO ($J$=1--0), (b) $^{13}$CO ($J$=1--0), and (c)  C$^{18}$O ($J$=1--0) in units of km\,s$^{-1}$ with effective beam sizes of 28$\farcs$3, 28$\farcs$6, and 28$\farcs$6, respectively. The gray and black dashed lines indicate in (b) the narrow filamentary structure and the redshifted streamer, respectively. The velocity dispersion maps of (d) $^{12}$CO ($J$=1--0), (e) $^{13}$CO ($J$=1--0), and (f)  C$^{18}$O ($J$=1--0) in units of km\,s$^{-1}$ with effective beam sizes of 28$\farcs$3, 28$\farcs$6, and 28$\farcs$6, respectively. The mean velocities and velocity dispersions of $^{12}$CO, $^{13}$CO, and C$^{18}$O maps are calculated in $v_{\rm LSR}$ = [$-$15\,km\,s$^{-1}$, 20\,km\,s$^{-1}$], [3\,km\,s$^{-1}$, 8\,km\,s$^{-1}$], and [5.7\,km\,s$^{-1}$, 7.5\,km\,s$^{-1}$], respectively. The rest symbols in the panels have the same meaning as in Fig.\,1.}
	\label{fig_mom}
\end{figure}
\clearpage

\setcounter{figure}{0}
\renewcommand{\thefigure}{B\arabic{figure}}

\begin{figure}
	\includegraphics[scale=.7]{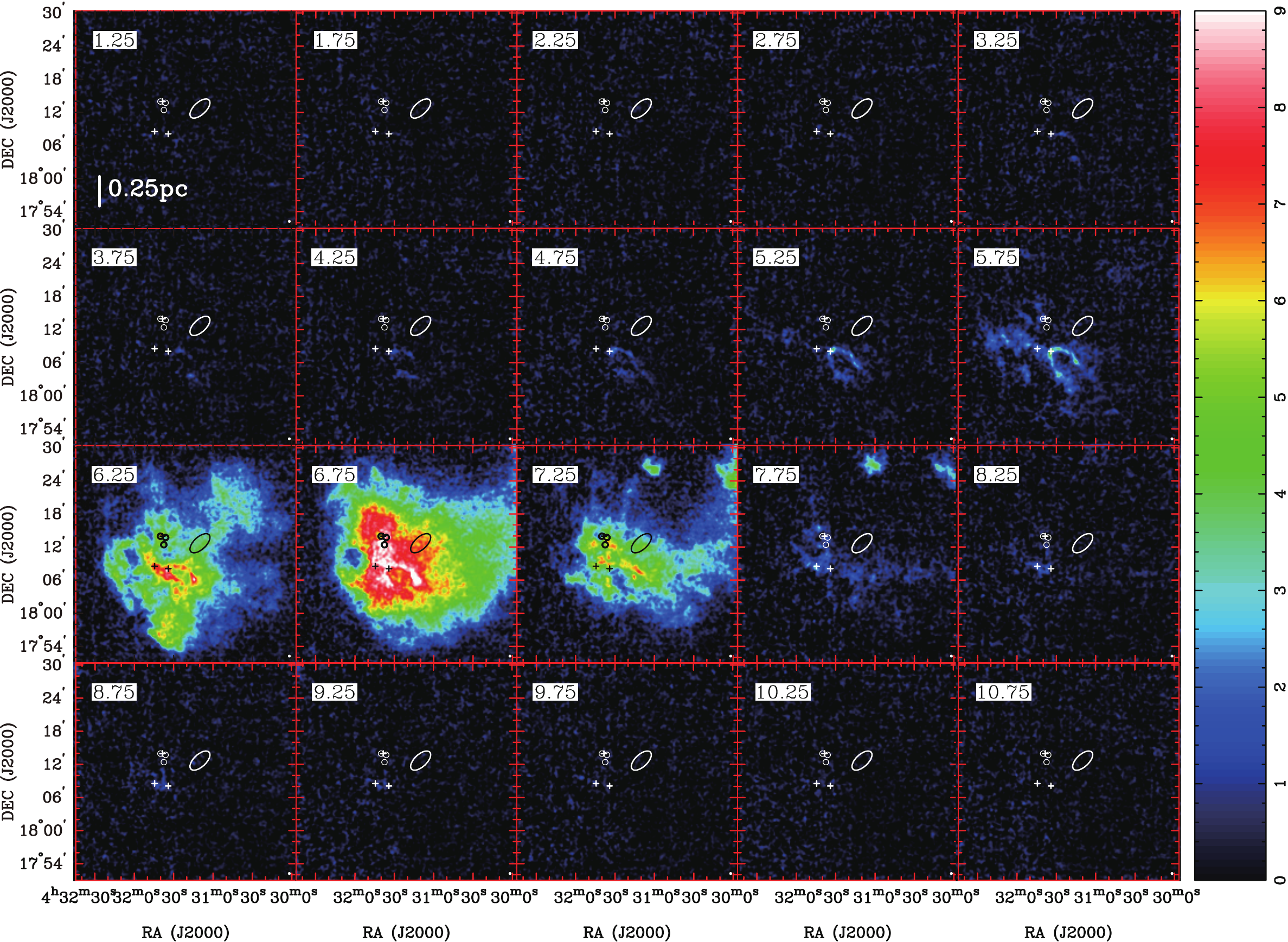}
	\caption{L\,1551 velocity channel maps of $^{13}$CO ($J$=1--0) in units of K. The numbers at the top-left corners denote the central  LSR velocities of the individual channels in units of km\,s$^{-1}$. The effective beam sizes are 30$\farcs$3. The linear scale of 0.25\,pc is shown at the bottom-left corner of the first panel.\label{fig_ch_13CO}}
\end{figure}
\clearpage

\setcounter{figure}{0}
\renewcommand{\thefigure}{C\arabic{figure}}

\begin{figure}
	\includegraphics[scale=.7]{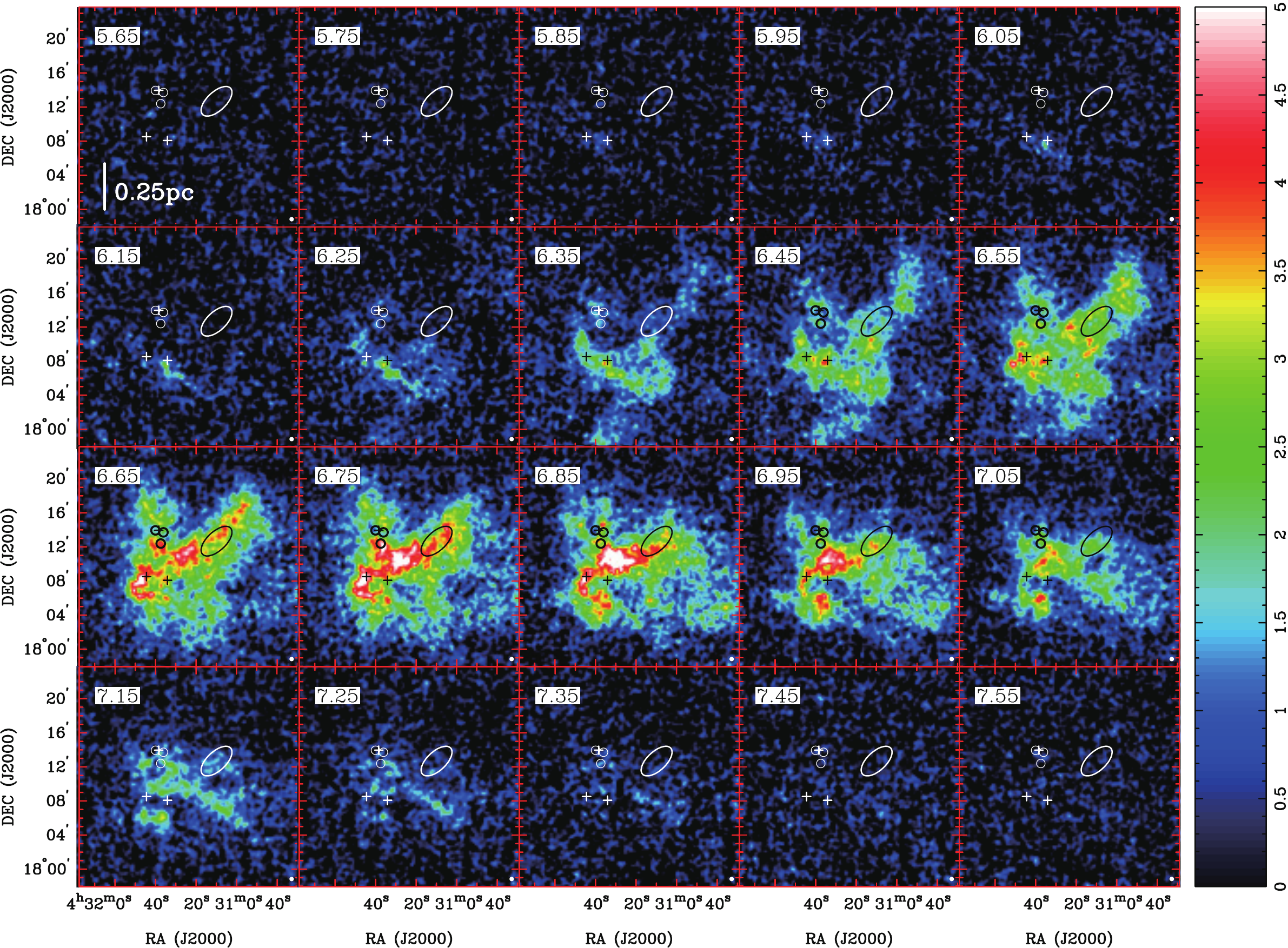}
	\caption{L\,1551 velocity channel maps of C$^{18}$O ($J$=1--0) in units of K. The numbers at the top-left corners denote the central  LSR velocities of the individual channels in units of km\,s$^{-1}$. The effective beam sizes are 30$\farcs$4. The linear scale of 0.25\,pc is shown at the bottom-left corner of the first panel.\label{fig_ch_C18O}}
\end{figure}
\clearpage

\setcounter{figure}{0}
\renewcommand{\thefigure}{D\arabic{figure}}

\begin{figure}
	\includegraphics[scale=.6]{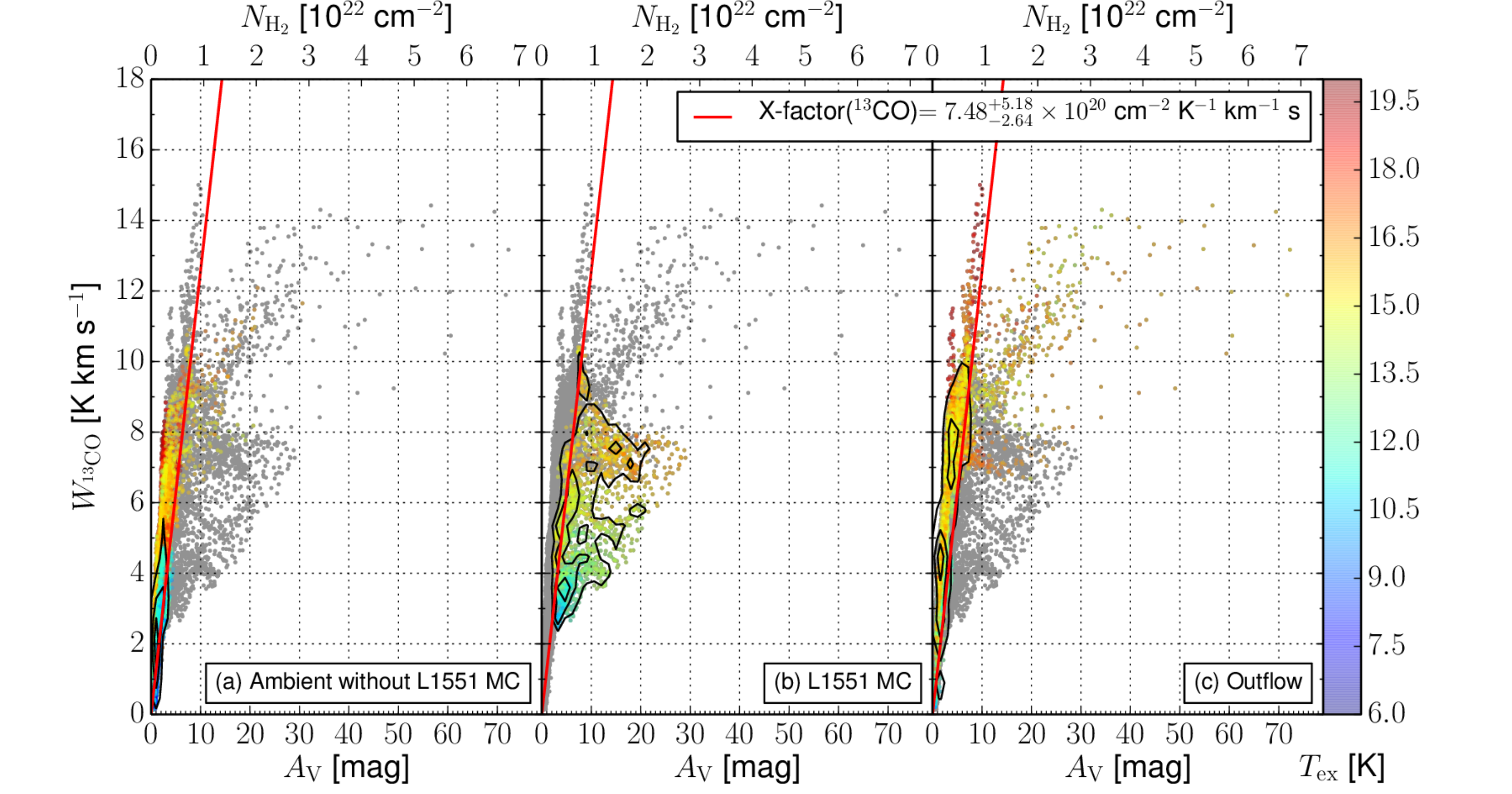}
	\caption{Correlation between the visual extinction, $A_{\rm V}$, and the integrated intensity of $^{13}$CO ($J$=1--0), $W_{\rm ^{13}CO}$. The gray dots denotes all data points in the map. The color-coded dots show the data points in the labeled regions, and the color represents the excitation temperature, $T_{\rm ex}$. The black contours indicate the surface density of the color-coded points at the 15$\%$, 45$\%$, and 75$\%$ levels of the maximum surface density in each panel.}
	\label{fig_Av_W13}
\end{figure}
\clearpage

\begin{figure}
	\includegraphics[scale=.6]{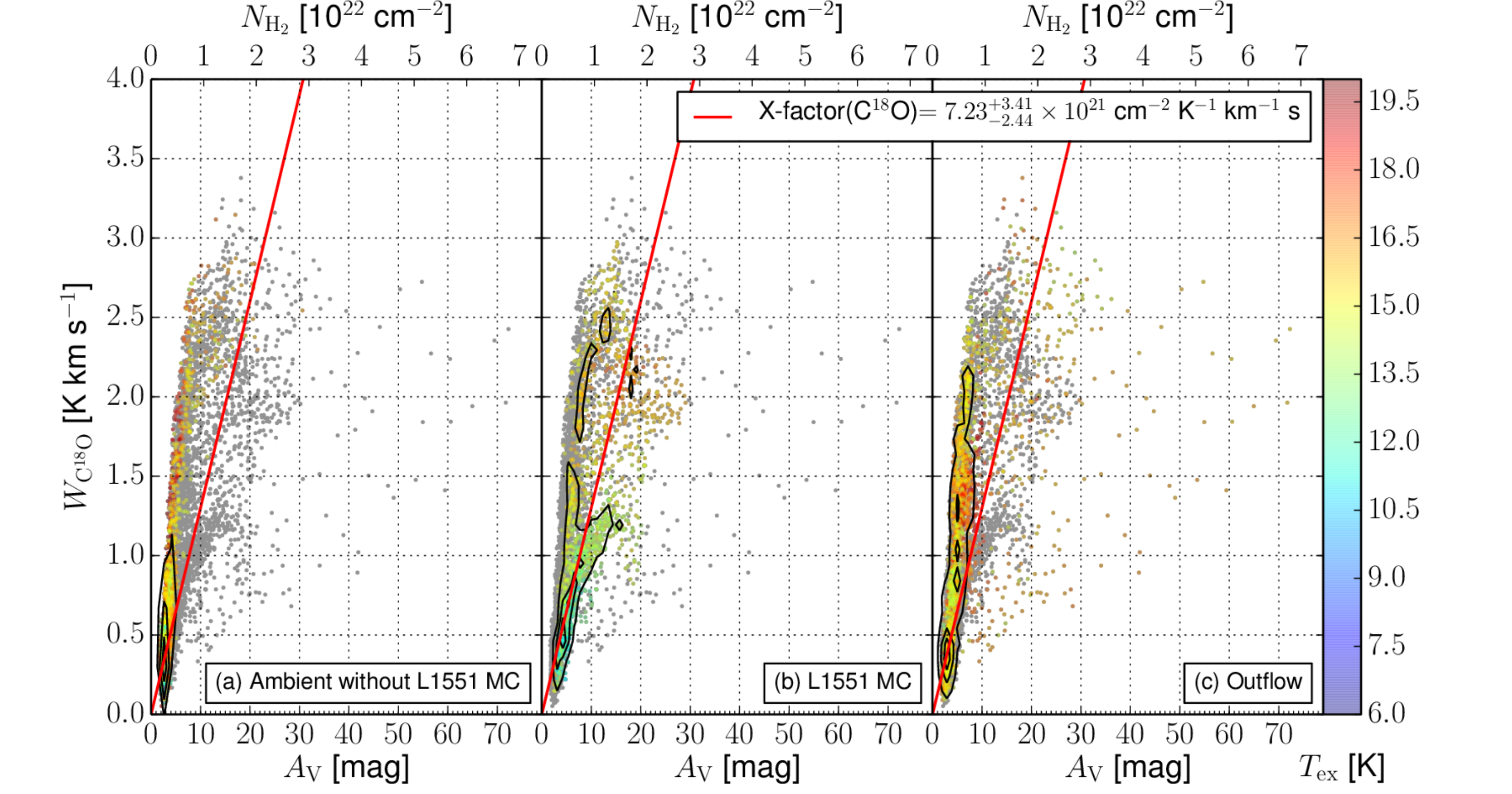}
	\caption{Correlation between the visual extinction, $A_{\rm V}$, and the integrated intensity of C$^{18}$O ($J$=1--0), $W_{\rm C^{18}O}$. The gray dots denotes all data points in the map. The color-coded dots show the data points in the labeled regions, and the color represents the excitation temperature, $T_{\rm ex}$. The black contours indicate the surface density of the color-coded points at the 15$\%$, 45$\%$, and 75$\%$ levels of the maximum surface density in each panel.}
	\label{fig_Av_W18}
\end{figure}
\clearpage

\setcounter{table}{0}

\begin{table}
	\caption{Parameters of our observations.\label{obs_parameter}}
	\begin{tabular}{lccc}
		\tableline\tableline
		Molecular line & $^{12}$CO ($J$=1--0) & $^{13}$CO ($J$=1--0) & C$^{18}$O ($J$=1--0)\\
		\tableline
		Rest Frequency [GHz] & 115.27120 & 110.20135 & 109.78218 \\
		Primary beam HPBW [$\arcsec$] & $\sim$15 & $\sim$16 & $\sim$16 \\
		Frequency resolution [kHz] & 37.8 & 37.8 & 37.8 \\
		Main beam efficiency & 0.32 & 0.38 & 0.43 \\ 
		Observation & 2007 Dec--2008 Jun & 2009 Dec--2010 Feb & 2009 Feb--2009 May\\
		Scan mode & OTF & OTF & OTF \\
		Mapping area [arcmin$^2$] & 44 $\times$ 44 & 42 $\times$ 43 & 30 $\times$ 30\\
		Pixel grid size [$\arcsec$] & 7.5 & 7.5 & 7.5\\
		Effective beam size [$\arcsec$] & 21.8 & 22.2 & 22.2\\  
		Velocity resolution [km\,s$^{-1}$] & 0.1 & 0.1 & 0.1\\
		Typical noise level in $T_{\rm mb}$ [K] & 1.23 & 0.94 & 0.67\\
		\tableline
	\end{tabular}
\end{table}


\begin{thebibliography}{}
	\bibitem[Alonso-Albi et al.(2010)]{alo10} Alonso-Albi, T., Fuente, A., Crimier, N., et al.\ 2010, \aap, 518, A52 
	\bibitem[Aniano et al.(2011)]{ani11} Aniano, G., Draine, B.~T., Gordon, K.~D., \& Sandstrom, K.\ 2011, \pasp, 123, 1218 
	\bibitem[Bergin \& Tafalla(2007)]{ber07} Bergin, E.~A., \& Tafalla, M.\ 2007, \araa, 45, 339 
	\bibitem[Bertout et al.(1999)]{ber99} Bertout, C., Robichon, N., \& Arenou, F.\ 1999, \aap, 352, 574 
	\bibitem[Bethell et al.(2007)]{bet07} Bethell, T.~J., Zweibel, E.~G., \& Li, P.~S.\ 2007, \apj, 667, 275 
	\bibitem[Bohlin et al.(1978)]{boh78} Bohlin, R.~C., Savage, B.~D., \& Drake, J.~F.\ 1978, \apj, 224, 132 
	\bibitem[Bolatto et al.(2013)]{bol13} Bolatto, A.~D., Wolfire, M., \& Leroy, A.~K.\ 2013, \araa, 51, 207 
	\bibitem[Burrows et al.(1996)]{bur96} Burrows, C.~J., Stapelfeldt, K.~R., Watson, A.~M., et al.\ 1996, \apj, 473, 437 
	\bibitem[Caselli et al.(1999)]{cas99} Caselli, P., Walmsley, C.~M., Tafalla, M., Dore, L., \& Myers, P.~C.\ 1999, \apjl, 523, L165 
	\bibitem[Devine et al.(1999)]{dev99} Devine, D., Reipurth, B., \& Bally, J.\ 1999, \aj, 118, 972 
	\bibitem[Dickman(1978)]{dic78} Dickman, R.~L.\ 1978, \apjs, 37, 407 
	\bibitem[Chou et al.(2014)]{cho14} Chou, T.-L., Takakuwa, S., Yen, H.-W., Ohashi, N., \& Ho, P.~T.~P.\ 2014, \apj, 796, 70 
	\bibitem[Clark 	\& Glover(2015)]{cla15} Clark, P.~C., \& Glover, S.~C.~O.\ 2015, \mnras, 452, 2057 
	\bibitem[Emerson et al.(1984)]{emr84} Emerson, J.~P., Harris, S., Jennings, R.~E., et al.\ 1984, \apjl, 278, L49 
	\bibitem[Emerson \& Graeve(1988)]{emr88} Emerson, D.~T., \& Graeve, R.\ 1988, \aap, 190, 353
	\bibitem[Ford \& Shirley(2011)]{for11} Ford, A.~B., \& Shirley, Y.~L.\ 2011, \apj, 728, 144 
	\bibitem[Frerking et al.(1982)]{fre82} Frerking, M.~A., Langer, W.~D., \& Wilson, R.~W.\ 1982, \apj, 262, 590 
	\bibitem[Griffin et al.(2010)]{gri10} Griffin, M.~J., Abergel, A., Abreu, A., et al.\ 2010, \aap, 518, LL3 
	\bibitem[Gutermuth et al.(2009)]{gut09} Gutermuth, R.~A., Megeath, S.~T., Myers, P.~C., et al.\ 2009, \apjs, 184, 18 
	\bibitem[Herbig et al.(2004)]{her04} Herbig, G.~H., Andrews, S.~M., \& Dahm, S.~E.\ 2004, \aj, 128, 1233 
	\bibitem[Hirota et al.(2008)]{hir08} Hirota, T., Ando, K., Bushimata, T., et al.\ 2008, \pasj, 60, 961 
	\bibitem[Hollenbach \& Tielens(1997)]{hol97} Hollenbach, D.~J., \& Tielens, A.~G.~G.~M.\ 1997, \araa, 35, 179 
	\bibitem[Kawamura et al.(1998)]{kaw98} Kawamura, A., Onishi, T., Yonekura, Y., et al.\ 1998, \apjs, 117, 387
	\bibitem[K{\"o}nyves et al.(2010)]{kon10} K{\"o}nyves, V., Andr{\'e}, P., Men'shchikov, A., et al.\ 2010, \aap, 518, LL106 
	\bibitem[Kong et al.(2015)]{kon15} Kong, S., Lada, C.~J., 
	Lada, E.~A., et al.\ 2015, \apj, 805, 58 
	\bibitem[Lada et al.(1994)]{lad94} Lada, C.~J., Lada, E.~A., Clemens, D.~P., \& Bally, J.\ 1994, \apj, 429, 694 
	\bibitem[Lada et al.(2009)]{lad09} Lada, C.~J., Lombardi, M., \& Alves, J.~F.\ 2009, \apj, 703, 52 
	\bibitem[Lee et al.(2014)]{lee14} Lee, M.-Y., Stanimirovi{\'c}, S., Wolfire, M.~G., et al.\ 2014, \apj, 784, 80 
	\bibitem[Liszt(2007)]{lis07} Liszt, H.~S.\ 2007, \aap, 476, 291 
	\bibitem[Looney et al.(1997)]{loo97} Looney, L.~W., Mundy, L.~G., \& Welch, W.~J.\ 1997, \apjl, 484, L157
	\bibitem[Menten et al.(2007)]{men07} Menten, K.~M., Reid, M.~J., Forbrich, J., \& Brunthaler, A.\ 2007, \aap, 474, 515 
	\bibitem[Moriarty-Schieven \& Wannier(1991)]{mor91} Moriarty-Schieven, G.~H., \& Wannier, P.~G.\ 1991, \apjl, 373, L23 
	\bibitem[Moriarty-Schieven et al.(2006)]{mor06} Moriarty-Schieven, G.~H., Johnstone, D., Bally, J., \& Jenness, T.\ 2006, \apj, 645, 357 
	\bibitem[Mundt et al.(1990)]{mun90} Mundt, R., Buehrke, T., Solf, J., Ray, T.~P., \& Raga, A.~C.\ 1990, \aap, 232, 37 
	\bibitem[Pineda et al.(2010)]{pin10} Pineda, J.~L., Goldsmith, P.~F., Chapman, N., et al.\ 2010, \apj, 721, 686 
	\bibitem[Poglitsch et al.(2010)]{pog10} Poglitsch, A., Waelkens, C., Geis, N., et al.\ 2010, \aap, 518, LL2 
	\bibitem[Pound \& Bally(1991)]{pou91} Pound, M.~W., \& Bally, J.\ 1991, \apj, 383, 705 
	\bibitem[Press et al.(2007)]{pre07} Press, W.~H., Teukolsky, S.~A., Vetterling, W.~T., \& Flannery, B.~P. 2007,	Numerical Recipes (Cambridge: Cambridge Univ. Press), 778
	\bibitem[Qian et al.(2012)]{qia12} Qian, L., Li, D., \& Goldsmith, P.~F.\ 2012, \apj, 760, 147 	
	\bibitem[Reipurth et al.(2002)]{rei02} Reipurth, B., Rodr{\'{\i}}guez, L.~F., Anglada, G., \& Bally, J.\ 2002, \aj, 124, 1045 
	\bibitem[R{\"o}llig \& Ossenkopf(2013)]{rol13} R{\"o}llig, M., \& Ossenkopf, V.\ 2013, \aap, 550, AA56 
	\bibitem[Sandstrom et al.(2007)]{san07} Sandstrom, K.~M., Peek, J.~E.~G., Bower, G.~C., Bolatto, A.~D., \& Plambeck, R.~L.\ 2007, \apj, 667, 1161 
	\bibitem[Sawada et al.(2008)]{saw08} Sawada, T., Ikeda, N., Sunada, K., et al.\ 2008, \pasj, 60, 445 
	\bibitem[Schwab(1984)]{sch84} Schwab, F. R. 1984, in Optimal Gridding of Visibility Data in Radio Interfer- ometry, Indirect Imaging, ed. J. A. Robert (Cambridge: Cambridge Univ. Press), 333
	\bibitem[Shetty et al.(2011a)]{she11a} Shetty, R., Glover, S.~C., Dullemond, C.~P., \& Klessen, R.~S.\ 2011, \mnras, 412, 1686 
	\bibitem[Shetty et al.(2011b)]{she11b} Shetty, R., Glover, S.~C., Dullemond, C.~P., et al.\ 2011, \mnras, 415, 3253 
	\bibitem[Shimajiri et al.(2011)]{shi11} Shimajiri, Y., Kawabe, R., Takakuwa, S., et al.\ 2011, \pasj, 63, 105 
	\bibitem[Shimajiri et al.(2013)]{shi13} Shimajiri, Y., Sakai, T., Tsukagoshi, T., et al.\ 2013, \apjl, 774, LL20 
	\bibitem[Shimajiri et al.(2014)]{shi14} Shimajiri, Y., Kitamura, Y., Saito, M., et al.\ 2014, \aap, 564, AA68
	\bibitem[Shimajiri et al.(2015)]{shi15} Shimajiri, Y., Kitamura, Y., Nakamura, F., et al.\ 2015, \apjs, 217, 7 
	\bibitem[Snell et al.(1980)]{sne80} Snell, R.~L., Loren, R.~B., \& Plambeck, R.~L.\ 1980, \apjl, 239, L17 
	\bibitem[Snell(1981)]{sne81} Snell, R.~L.\ 1981, \apjs, 45, 121 
	\bibitem[Sorai et al.(2000)]{sor00} Sorai, K., Sunada, K., Okumura, S. K., et al. 2000, in Society of Photo-Optical Instrumentation Engineers (SPIE) Conference Series, Vol. 4015, Radio Telescopes, ed. H. R. Butcher, 86-95
	\bibitem[Stojimirovi{\'c} et al.(2006)]{sto06} Stojimirovi{\'c}, I., Narayanan, G., Snell, R.~L., \& Bally, J.\ 2006, \apj, 649, 280
	\bibitem[Swift et al.(2005)]{swi05} Swift, J.~J., Welch, W.~J., \& Di Francesco, J.\ 2005, \apj, 620, 823
	\bibitem[Swift et al.(2006)]{swi06} Swift, J.~J., Welch, W.~J., Di Francesco, J., \& Stojimirovi{\'c}, I.\ 2006, \apj, 637, 392  
	\bibitem[Swift \& Welch(2008)]{swi08} Swift, J.~J., \& Welch, W.~J.\ 2008, \apjs, 174, 202 
	\bibitem[Sunada et al.(2000)]{sun00} Sunada, K., Yamaguchi, C., Nakai, N., et al. 2000, in Society of Photo-Optical Instrumentation Engineers (SPIE) Conference Series, Vol. 4015, Radio Telescopes, ed. H. R. Butcher, 237-246
	\bibitem[Sz{\H u}cs et al.(2014)]{szu14} Sz{\H u}cs, L., Glover, S.~C.~O., \& Klessen, R.~S.\ 2014, \mnras, 445, 4055 
	\bibitem[Tachihara et al.(2002)]{tac02} Tachihara, K., Onishi, T., Mizuno, A., \& Fukui, Y.\ 2002, \aap, 385, 909 
	\bibitem[Takakuwa et al.(2014)]{tak14} Takakuwa, S., Saito, M., Saigo, K., et al.\ 2014, \apj, 796, 1 
	\bibitem[Ulich \& Haas(1976)]{uli76} Ulich, B.~L., \& Haas, R.~W.\ 1976, \apjs, 30, 247 
	\bibitem[van Dishoeck \& Black(1988)]{dis88} van Dishoeck, E.~F., \& Black, J.~H.\ 1988, \apj, 334, 771 
	\bibitem[Visser et al.(2009)]{vis09} Visser, R., van Dishoeck, E.~F., \& Black, J.~H.\ 2009, \aap, 503, 323 
	\bibitem[Warin et al.(1996)]{war96} Warin, S., Benayoun, J.~J., \& Viala, Y.~P.\ 1996, \aap, 308, 535 
	\bibitem[Wilson(1999)]{wil99} Wilson, T.~L.\ 1999, Reports on Progress in Physics, 62, 143
	\bibitem[Yoshida et al.(2010)]{yos10} Yoshida, A., Kitamura, Y., Shimajiri, Y., \& Kawabe, R.\ 2010, \apj, 718, 1019
 
\end{thebibliography}
\end{document}